**Chemical reaction directed oriented attachment: from precursor particles to new substances**

Yongfei Liu[a], Xiaoying Qin[a], Yong Yang[a], Zhi Zeng[a], Shuangming Chen[b], Yunxiang Lin[b], Hongxing Xin[a], Zhengfei Dai[a], Chunjun Song[a], Xiaoguang Zhu[a], Di Li[a], Jian Zhang[a], Li Song[b] and Yoshiyuki Kawazoe[c,d]

[a]*Key Laboratory of Materials Physics, Institute of Solid State Physics, Chinese Academy of Sciences, Hefei 230031, China.*

[b]*National Synchrotron Radiation Laboratory, CAS Center for Excellence in Nanoscience, University of Science and Technology of China, Hefei, Anhui 230029, China.*

[c]*New Industry Creation Hatchery Center (NICHe), Tohoku University, 6-6-4 Aoba, Aramaki, Aoba-ku, Sendai, Miyagi 980-8579, Japan.*

[d]*Institute of Thermophysics, Siberian Branch of Russian Academy of Sciences, Novosibirsk 630090, Russia.*

**ABSTRACT:** The oriented attachment (OA) of nanoparticles is an important mechanism for the synthesis of the crystals of inorganic functional materials, and the formation of natural minerals. For years it has been generally acknowledged that OA is a physical process, i.e., particle alignments and interface fusion via mass diffusion, not involving the formation of new substances. Hence, the obtained crystals maintain identical crystallographic structures and chemical constituents to those of the precursor particles. Here we report a chemical reaction directed OA growth, through which $Y_2(CO_3)_3 \cdot 2H_2O$ nanoparticles are converted to single-crystalline double-carbonates (e.g., $NaY(CO_3)_2 \cdot 6H_2O$). The dominant role of OA growth is supported by our first-principles calculations. Such a new OA mechanism enriches the aggregation-based crystal growth theory.

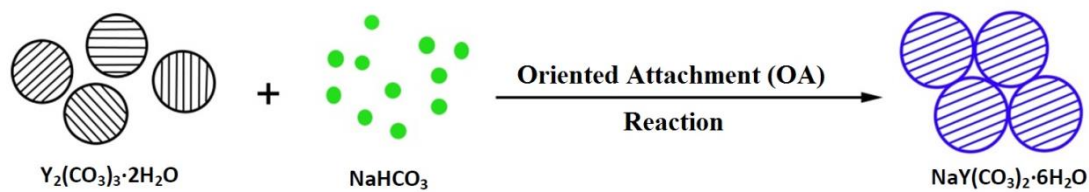

**INTRODUCTION**

As a nonclassical crystal growth mode, oriented attachment (OA) plays an increasingly important role in materials science and nanotechnology, and has shown its significance in the development of the theory of crystal growth and mineralogy.[1-3] The OA process usually involves the oriented self-assembly of primary nanoparticles and conversion to single crystals or pseudocrystals by interface fusion (**Figure 1A**).[2, 3] It is theorized that the OA process is dominant in the early stage of crystal growth,[4] driven by the Brownian motion or short-range interactions.[2,5] Such oriented alignments usually can be achieved under hydrothermal/solvothermal conditions[3, 6, 7] or with the assistance of surfactants (**Figure 1B**).[8-11] In conventional OA growth, the constituent and phase structure of the formed crystals are identical to those of the precursor particles. Here we report a chemical reaction directed OA growth that can create new substances from the precursor nanoparticles (**Figure 1C**). Specifically, we show that through such a new OA mechanism, $Y_2(CO_3)_3 \cdot 2H_2O$ nanoparticles can grow into single-crystalline sandwich-structured $NaY(CO_3)_2 \cdot 6H_2O$ and $(NH_4)Y(CO_3)_2 \cdot H_2O$ microsheets by reacting with $NaHCO_3$ and $NH_4HCO_3$, respectively. The assembled architectures are characterized by layer and spiral growth and other interesting features, such as kinks, terraces, and steps, indicating that during the OA process the nanoparticles behave as ions or molecules do in the classical mode of crystal growth. Such a chemical reaction directed OA growth can be applied to the synthesis of other alkali-metal rare-earth double carbonates. Our first-principles calculations demonstrate that the surface adsorption induced reaction plays a dominant role in the formation of the yttrium double carbonates. Present finding not only enriches the aggregation-based crystal growth theory, but also paves a new way for designing advanced functional materials from the elementary nanoparticles.

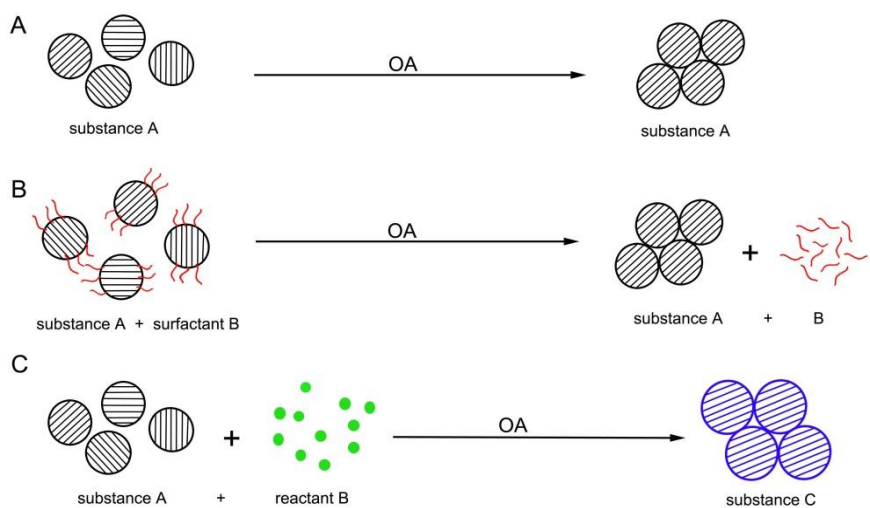

**Figure 1. Schematics of conventional OA processes** (*A* and *B*) **and chemical reaction directed OA process** (*C*). (*A*) The oriented alignments of primary nanoparticles and fusion to form a single crystal. (*B*) With the assistance of surfactants (red wavy lines) a single crystal grows via oriented

alignments and fusion of primary nanoparticles. (*C*) New single-crystalline substance forms accompanying the alignment and reorganization of precursor nanoparticles while reacting with introduced reactants.

**OA growth of precursor nanoparticles into new compounds.** In a typical synthesis protocol, the precursor $Y_2(CO_3)_3$ $2H_2O$ nanoparticles (labeled as Y0) were fabricated via a conventional precipitation method.[12] To the suspensions, $NaHCO_3$ and $NH_4HCO_3$ aqueous solutions at the desired concentration were added as reactants under vigorous stirring, respectively. After reaction for 10-120 min, single-crystalline double-carbonates of $NaY(CO_3)_2$ $6H_2O$ and $(NH_4)Y(CO_3)_2$ $H_2O$ sheets of micrometer sizes (labeled as Y1 and Y2, respectively) were successfully synthesized. Details of the synthesis procedure can be found in the Methods section.

## RESULTS AND DISCUSSION

We performed transmission electron microscopy (TEM), field-emission scanning electron microscopy (FESEM), atomic force microscopy (AFM), powder X-ray diffraction (PXRD), and small angle X-ray scattering (SAXS) analyses on the precursor nanoparticles and the synesized sheet-like double-carbonates. The TEM image of the precursor (**Figure 2A**) shows that it is composed of weakly flocculated nanoparticles with sizes ranging from ~ 10 to 60 nm, in good agreement with SAXS measurements (**Figure 2E**). In contrast, the obtained sheets possess rectangular and rhombic shapes in lateral sizes up to ~ 10 μm (**Figures 2B-C**, **Figure S1**). The selected-area electron diffraction (SAED) patterns (insets of **Figures 2B**-**C**) verify that these micron-sized sheets are single crystals. The PXRD patterns (**Figure 2F**) confirm that the precursor nanoparticles and the obtained sheets are $Y_2(CO_3)_3$ $2H_2O$ (orthorhombic, JCPDS card No. 81-1538), $NaY(CO_3)_2$ $6H_2O$ (anorthic, JCPDS card No. 54-0703), and $(NH_4)Y(CO_3)_2$ $H_2O$ [see **Supporting Information** Text], respectively. This result indicates that both chemical constituents and crystal structures have changed after the growth, leading to the formation of new substances.

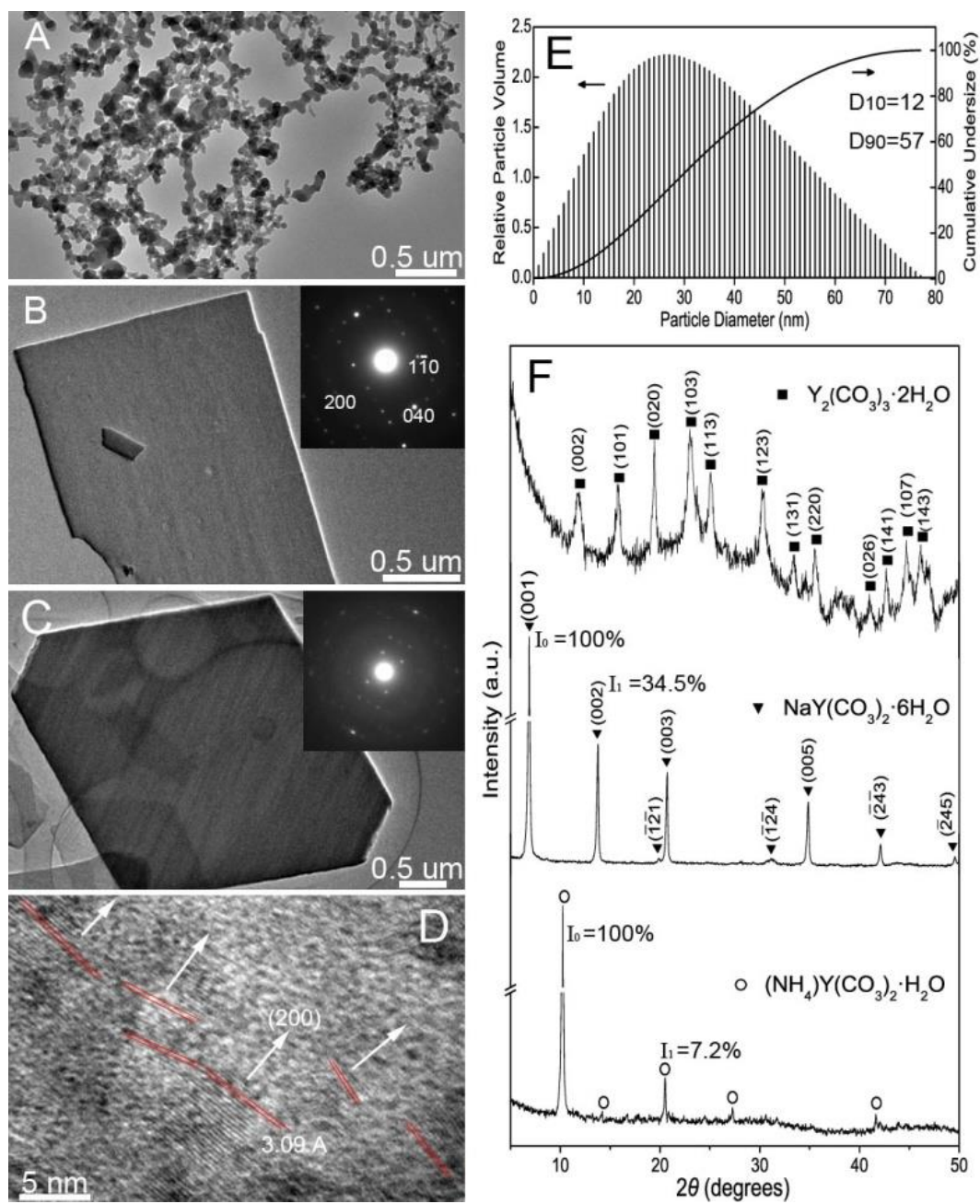

**Figure 2. Characterization of the precursors and products.** (*A-C*) TEM images of precursor nanoparticles (*A*) and the two obtained sheets Y1 (*B*) and Y2 (*C*). Insets are corresponding SAED patterns. (*D*) HRTEM image of Y1 sheet. Imperfect oriented attachments can be identified as indicated by the red lines. (*E*) Particle size distribution of the precursor nanoparticles measured by SAXS. (*F*) PXRD patterns of the precursor nanoparticles and obtained sheets, corresponding to orthorhombic $Y_2(CO_3)_3 \cdot 2H_2O$, anorthic $NaY(CO_3)_2 \cdot 6H_2O$ and $(NH_4)Y(CO_3)_2 \cdot H_2O$ phases, respectively.

Our high resolution TEM (HRTEM) images show that many regular lattice domains (**Figure 2D** and **Figure S2A**) in an average size of ca. 5 nm exist in the grown $NaY(CO_3)_2 \cdot 6H_2O$ single crystal, which is hardly understood by the classical ion-by-ion growth mode. The sizes of these domains are comparable to the primary particles of the precursors (**Figure S2B**). Especially, as shown in the **Figure S2C**, one can see that the domain sizes of the precursors and Y1 sheets are in the ranges from ~1.8 to 6.2 nm and from ~2.7 to 7.9 nm, respectively, indicating good agreements in their domain sizes. Such aligned domain structures and the correspondence between the domains in the grown crystal and the primary particles of precursors strongly indicate an oriented attachment of nanoparticles. A slight increase in the domain size should be due to the diffusion of $Na^+$ ($NH_4^+$) and $CO_3^{2-}$ into the lattice of the precursors (see discussions below), and the perfect crystallographic reorganization of the domains within the self-assemblies. Though the clear lattice-fringe image of $(NH_4)Y(CO_3)_2 \cdot H_2O$ sheet cannot be obtained due to the rapid beam damage in TEM observations (**Figure S3**), the sharp XRD peaks (at $2\theta=6.88°$ for $NaY(CO_3)_2 \cdot 6H_2O$ and $10.21°$ for $(NH_4)Y(CO_3)_2 \cdot H_2O$) suggest their high crystallinity and oriented growth.[13]

**Granular nature of the fracture surfaces for the grown sheets**. As revealed by the FESEM inspections, the two sheets show parallel edges, spiral patterns (**Figures. 3A-B**, and **Figure S1**), and other surface defects such as steps, terraces and kinks (indicated by arrows in **Figure 3A**), which are reminiscent of the classical crystal growth that involves the ion-by-ion attachment.[14] Nevertheless, detailed FESEM/AFM observations show that the fracture surfaces of the sheets have granular fracture nature (indicated by red arrow 3 in **Figure 3C**, **Figures 3E-F**), indicating that they consist of nanoparticle assemblies. As revealed by the size distribution of the exposed particles, these fracture surfaces display particles that have similar dimension (~ 10 to 60 nm, **Figure S4**) to those of the secondary nanoparticles of the precursor $Y_2(CO_3)_3 \cdot 2H_2O$ (as shown in **Figures 2A & 2E**). One can also find this particle-assembly characteristic from the vertical plane of the fracture (indicated by red arrow 4 in **Figure 3D**). Such fracture morphologies can never appear in the single crystal grown through the classical mode, which will exhibit cleavage fractures being composed of atomically flat cleavage planes (especially for brittle materials such as salts). In contrast, during the OA process the surfaces of nanoparticles are not atomically flat and the size of the nanoparticles is much larger than atoms/ions. As a result, imperfect alignments of nanoparticles and incomplete interface fusion occur frequently, which inevitably results in the formation of defect-rich regions at the interfaces of the precursor nanoparticles.[3] While the OA grown single crystals fracture, the cracks will propagate along these defect-rich interface regions and the fracture surfaces will show granular feature, and thus exposing characteristics of nanoparticle assemblies. No such features are observed on the Si substrates (indicated by green arrows 1 and 2 in **Figures 3B and 3D**), even on the unbroken surfaces of the sheets (blue arrow 5 in **Figure 3D**), indicating that these granular fractures do not come from the deposition of nanoparticles but originate from the inherent nature of the sheets. Hence, present observations further verify that the single crystalline sheets originate from the oriented attachment of nanoparticles.

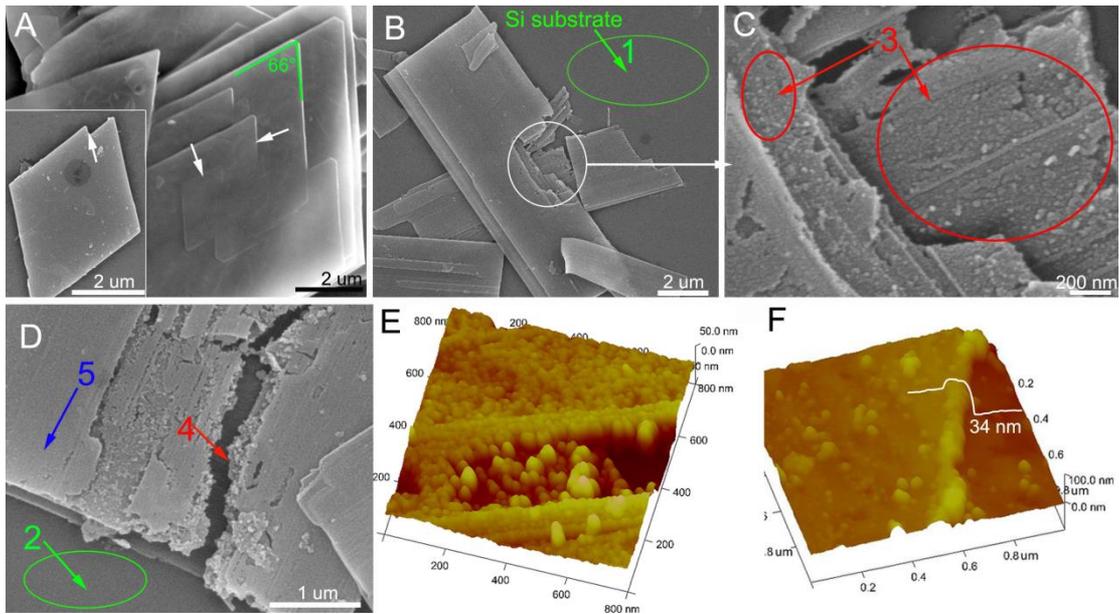

**Figure 3. Surface and fracture surface morphologies of Y1 and Y2 sheets.** (*A-B*), FESEM images showing surfaces morphologies of (*A*) Y2 and (*B*) Y1 sheets. (*C-D*) FESEM images showing morphologies of fracture surfaces of Y1 sheets. (*E*) 3D AFM image showing fracture surface morphology of Y1 sheet. (*F*), 3D AFM surface plot Y2 sheets. The inset line in (*F*) shows the cross-sectional analysis of the step site.

**The microstructures of the intermediates.** We further study the formation of intermediates for NaY(CO$_3$)$_2$·6H$_2$O sheets taken from the suspension synthesized at 0.2 M of Y(NO$_3$)$_3$ solution and 1 M NaHCO$_3$ solution after reaction for 1 hr. One can see that on the sheets there are some "unmatured" areas (**Figure S5**), which show network-like structure consisting of aggregated particles. SAED pattern shows the single crystalline nature of the matured sheet (**Figure S5C**), while the HRTEM images of the networks indicate that they consist of particles with disordered orientations (**Figure S5D**). Such features exclude the possibility that this kind of networks come from the partial dissolution of the pristine single crystalline sheets; for if they were, the networks should still show aligned lattice fringes rather than random ones. The particle-stacking-like structures are the intermediates between the starting particles and the final sheets. On the other hand, the SEM and TEM images for the intermediate products sampled during the formation of (NH$_4$)·Y(CO$_3$)$_2$·H$_2$O sheets (**Figure S6**) also indicate that the sheets are also grown through particle aggregation (i.e., OA mechanism) instead of the classical mode.

**The effect of the collision frequency on the OA process.** We studied the dependence of the OA growth on the effective collisions of NPs by diluting the initial Y0 suspensions and NaHCO$_3$ /NH$_4$HCO$_3$ solutions synchronously in the contrast experiments. As a result, no microsheets can be found in the either experiment (**Figures S7A** and **S7E**) as diluted to 1/50; while single-layer sheets begin to form (**Figures 7B** and **7F**) until the dilution rate reaches 1/10 (which means a higher concentration of NPs), indicating the key role of the collision frequency. Especially, we notice that single-layer Y2 sheets result from lower NPs concentration and multilayer spiral sheets are formed when higher concentrations are employed (**Figures S7F** and **S7G**). Such phenomenon is contrary to the classical crystal growth mechanism; therein spiral growth is dominant at the low concentration.[14] Meanwhile, intersecting Y2 sheets (**Figure S7H**) were obtained under the highest concentration Y0 suspension, which should arise from the mismatched attachment at the ultra-high NPs concentration.

**Particle assembly in the OA grown sheets revealed by etching-treatment of defect-rich regions.** Generally, OA and Ostwald ripening (OR) mechanisms coexist during the crystal growth,[4] and the latter can promote the elimination of grain boundaries and surface reconstruction.[15,16] However, as mentioned above, defect-rich regions at the interfaces of the precursor nanoparticles exist in the OA grown crystals and these regions can be etched preferentially by chemical etching due to their higher energy. In order to uncover these defect-rich regions in the smooth $(NH_4)Y(CO_3)_2 \cdot H_2O$ sheets, hydro/solvothermal treatments were conducted at temperature of 160 ℃ for 24 hrs in ethanol, ionized water and toluene, respectively. The etching reaction was tuned by controlling the addition of $NH_3 \cdot H_2O$, which can regulate the concentration of $NH_4^+$ and thus ensures that the defect-rich regions can be etched preferentially (see **Figure S8** and **Figure S9** for details). **Figures 4A-C** show the FESEM images of the sheets after being treated in ethanol. It can be clearly seen that the etched sheets are composed of regular particle array patterns in which the particle arrangements are highly ordered and have the same shape as the pristine rhombic sheets. Similarly, the sheets show rhombic patterns to which some scattered nanoparticles cling after being treated in both toluene (**Figures 4D-E**) and ionized water (**Figure 4F**). The acute angles of the rhombic frameworks are about 65°~68°, matching that of the pristine sheets (66°, as shown in **Figure 3A**). Such regular particle patterns cannot be obtained by etching a single crystal grown from the classical mode. Present observations reveal the intrinsic microstructures of pristine sheets, indicating the nanoparticle aggregation-based crystal growth. It should be noted that the average particle size (55 nm) in the crystal after etching treatment is about twice that (32 nm) of the original precursor nanoparticles (**Figure S9E**). This indicates that the exposed particles of the Y2 sheets after etching treatment are actually agglomerates of the precursor particles (each corresponds to approximately five precursor particles since $(55/32)^3 \approx 5$). Since etching takes place preferentially in defect-rich interfaces in-between particles, the exposed particles (agglomerates) of the Y2 sheets after etching are composed of well-aligned assembled particles with relatively perfect interface fusion. Hence, these particles exposed in etching treatment can be regarded as secondary particles participating in the OA process.

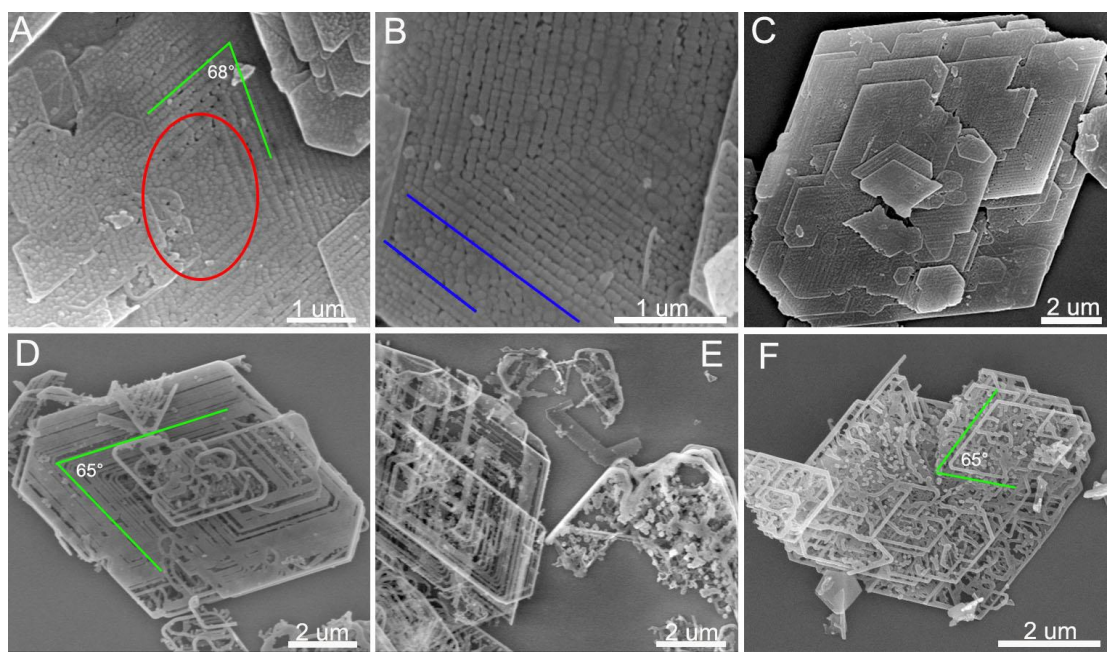

**Figure 4.** FESEM images of the $(NH_4)Y(CO_3)_2 \cdot H_2O$ sheets after post-treatment in: (*A-C*) ethanol, (*D, E*) toluene, and (*F*) ionized water, respectively. Typically, $(NH_4)Y(CO_3)_2 \cdot H_2O$ suspensions (3 ml) were mixed with 3 ml $NH_4OH$ solution (1.0 M) into 45 ml of solvents (ionized water, absolute ethyl alcohol, and toluene, respectively), and then aged at 160 ℃ for 24 hours.

**The role of the chemical reaction in the OA process.** It is obvious that chemical reactions between $Y_2(CO_3)_3 \cdot 2H_2O$ nanoparticles and $CO_3^{2-}$, $Na^+$, or $NH_4^+$ play a decisive role in the formation of single crystalline $NaY(CO_3)_2 \cdot 6H_2O$ (or $(NH_4)Y(CO_3)_2 \cdot H_2O$) sheets through the OA of precursor nanoparticles $(Y_2(CO_3)_3 \cdot 2H_2O)$. To examine the effects of the chemical reaction on the OA growth, the introduced reactants of $NaHCO_3$ and $NH_4HCO_3$ solutions were replaced by NaOH and $NH_4OH$ solutions (with the same concentration), respectively. FESEM observations show that no micron-sized or nano-sized sheets are obtained (**Figure S10**). This result indicates that the chemical reaction of the introduced reactants with $Y_2(CO_3)_3 \cdot 2H_2O$ to form double-salts are prerequisite to such an OA growth, since no double carbonates are produced through the addition of NaOH or $NH_4OH$. Moreover, we have successfully applied this strategy to synthesize other alkali-metal rare-earth double carbonates, such as $NaGd(CO_3)_2 \cdot 6H_2O$ and $KNd(CO_3)_2 \cdot xH_2O$. Single crystalline sheets can also be obtained through the reactions of $Gd_2(CO_3)_3 \cdot 2H_2O$ (or $Nd_2(CO_3)_3 \cdot 2H_2O$) nanoparticles with $NaHCO_3$ (or $KHCO_3$), as shown in **Figure S11**, which further evidences the pivotal role of the reaction to form double-salts in the OA growth.

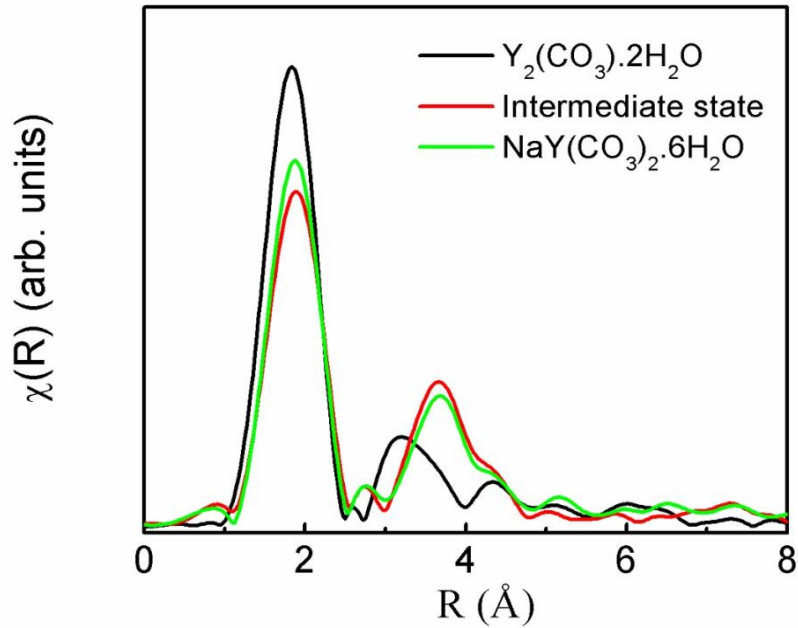

**Figure 5**. Radial structure functions (Fourier transform of the EXAFS oscillations) around the $Y^{3+}$ ions of Y0 suspension ($Y_2(CO_3)_3 \cdot 2H_2O$ nanoparticles), the final product Y2 suspension ($NaY(CO_3)_2 \cdot 6H_2O$), and the intermediate state (the suspensions sampled during the formation process of Y2). The intensity of the first coordination peak at about 1.9 Å is proportional to the oxygen coordination numbers around $Y^{3+}$ ions in the first atomic coordination shell.

Furthermore, we have also performed in-situ extended X-ray absorption fine structure (EXAFS) measurements to study the growth dynamics of the double carbonates, as shown in **Figure 5**. One can see that besides the height of the first peaks being different for Y0, Y2 suspension and the intermediate state, the height of the second peak also differs for Y2 suspension and the intermediate state. Moreover, the positions of the third and fourth peaks differ substantially for the three states, indicating the differences in their local atomic structures around $Y^{3+}$. This result demonstrates that there are changes in their crystallographic structures before and after OA process. This evidence rules out the possibility of conventional OA growth mode, for in a conventional OA growth mode the local atomic structures of $Y^{3+}$ ions in the nanoparticles do not vary before and after OA.

**The mechanism of the chemical reaction directed OA.** The most interesting issue is how the chemical reaction can give rise to the OA growth of the yttrium double carbonates ($NaY(CO_3)_2 \cdot 6H_2O$ and $(NH_4)Y(CO_3)_2 \cdot H_2O$). The layer structured $Y_2(CO_3)_3 \cdot 2H_2O$ consists of $YO_9$ polyhedra and the sandwich structured yttrium double carbonates consists of $YO_9$ polyhedra layers together with $Na(CO_3) \cdot H_2O$ or $NH_4 (CO_3) \cdot H_2O$ layers.[17, 18] Therefore, the reactions may take place through the approach of ion diffusions ($CO_3^{2-}$, $Na^+$ or $NH_4^+$) into the gallery space of $Y_2(CO_3)_3 \cdot 2H_2O$ lattice (**Figure S12**) due to the insolubility of $Y_2(CO_3)_3 \cdot 2H_2O$ ($K_{sp} = 1.03 \times 10^{-31}$ at 25 ℃),[19] as follows:

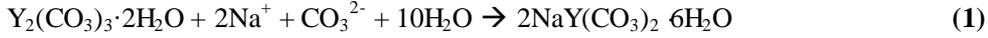

$$Y_2(CO_3)_3 \cdot 2H_2O + 2Na^+ + CO_3^{2-} + 10H_2O \rightarrow 2NaY(CO_3)_2 \cdot 6H_2O \qquad (1)$$

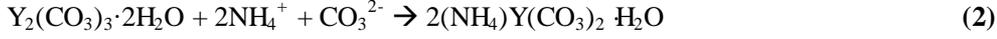

$$Y_2(CO_3)_3 \cdot 2H_2O + 2NH_4^+ + CO_3^{2-} \rightarrow 2(NH_4)Y(CO_3)_2 \cdot H_2O \qquad (2)$$

The large gallery space in the (100) plane (**Figure S12**) can serve as channels for the diffusions of the ions (the thermochemical radii of $CO_3^{2-}$, $Na^+$ and $NH_4^+$ are 1.78, 0.97, and 1.43 Å, respectively).[20, 21] On the other hand, the classical mode of crystal growth could also be possible. In the production of $NaY(CO_3)_2 \cdot 6H_2O$, the reactions may proceed as follows:

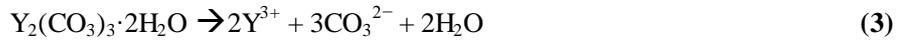

$$Y_2(CO_3)_3 \cdot 2H_2O \rightarrow 2Y^{3+} + 3CO_3^{2-} + 2H_2O \qquad (3)$$

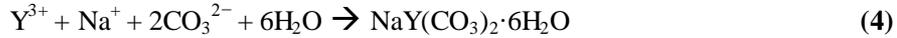

$$Y^{3+} + Na^+ + 2CO_3^{2-} + 6H_2O \rightarrow NaY(CO_3)_2 \cdot 6H_2O \qquad (4)$$

Similar processes are expected for the production of $(NH_4)Y(CO_3)_2 \cdot H_2O$. For classical mode of growth, due to the presence of excess $CO_3^{2-}$, the amount of aqueous $Y^{3+}$ will play a key role. Supposed that the aqueous $Y^{3+}$ and $CO_3^{2-}$ groups (where the initial concentration of $HCO_3^-$ is ~1/3 M in the mixed solution) are in equilibrium with the $Y_2(CO_3)_3 \cdot 2H_2O$ suspensions, then the concentration of $Y^{3+}$ is calculated to be ~ $4.1 \times 10^{-8}$ M (**Supporting Information** Text "Calculation of the Equilibrium Concentration of $Y^{3+}$"). Such a small amount of $Y^{3+}$ will be quickly consumed by the reaction (4). To maintain the solubility equilibrium, the Y ions of the $Y_2(CO_3)_3 \cdot 2H_2O$ suspensions are forced to diffuse into the solution, which is imperative for reaction (4) to proceed.

One can determine which process is predominant by comparing the corresponding rate constant $K$, as described by the Arrhenius equation: $K = Ae^{-E_a/(k_B T)}$, where $A$ is the frequency factor, $E_a$ is the activation energy, $k_B$ is the Boltzmann constant and $T$ is the temperature. For a given system, the activation energy $E_a$ plays a key role in the kinetics of chemical reaction. We have performed first-principles calculations to get the value of $E_a$ in each process. Details of the calculations are summarized in the section Methods. The obtained relative energies regarding the diffusion of $CO_3^{2-}$, $Na^+$, $NH_4^+$, and $Y_3^+$ along the [100] direction of $Y_2(CO_3)_3 \cdot 2H_2O$ crystal are shown in **Figure 6**, together with some intermediate atomic configurations (labeled by lowercase letters).

As shown in **Figure 6**, the diffusion of $CO_3^{2-}$ from surface (a) to the interior (c) of $Y_2(CO_3)_3 \cdot 2H_2O$ involves the surmounting of several barriers: Configurations a $\rightarrow$ b ($E_{a1} = 1.33$ eV); $u_1 \rightarrow u_2$ ($E_{a2} = 0.84$ eV); $u_3 \rightarrow u_4$ ($E_{a3} = 0.28$ eV). The other configuration transitions are spontaneous processes. Whereas, $Na^+$ can easily move into the $Y_2(CO_3)_3 \cdot 2H_2O$ crystal via the nanochannel at a depth of ~ 2.5 Å due to the small barrier ($E_{a4}$ ~ 0.18 eV, d $\rightarrow$ $w_2$). Then they face a high transition barrier ($w_2 \rightarrow e$, $E_{a5}$ ~ 1.36 eV) prior arriving at configuration f. The diffusion of $NH_4^+$ needs to overcome a first barrier $E_{a6}$ (~ 1.20 eV, g $\rightarrow$ $v_1$), and then a second $E_{a7}$ (~ 0.96 eV, $v_2 \rightarrow$ h), before reaching i. In contrast, the diffusion of $Y^{3+}$ from the interior (j) to the surface experiences a deep potential well at configuration k, and then needs to overcome a very high barrier $E_{a8}$ (~ 6.91 eV, k $\rightarrow$ l).

To a multistep diffusion, the averaged rate constant $K$ satisfies the relation: $\frac{1}{K} = \sum_{i=1}^{n} \frac{1}{K_i}$, where $K_i$ is the rate constant of step $i$ (see **Supporting Information** Text "Calculation of Rate Constant for a Multistep Diffusion" and **Figure S13**). With the calculated energy barriers, we can estimate the rate constant for the diffusion process by assuming that the frequency factor $A$ to be ~ $10^{12}$ s$^{-1}$.[22, 23] The values of $K$ are $8.3 \times 10^{-9}$ s$^{-1}$, $2.7 \times 10^{-9}$ s$^{-1}$, $6.9 \times 10^{-7}$ s$^{-1}$, and $3.1 \times 10^{-105}$ s$^{-1}$, for the diffusion of $CO_3^{2-}$, $Na^+$, $NH_4^+$ and $Y^{3+}$, respectively. Judging from the very small rate constant, it is unlikely that the diffusion of $Y^{3+}$ ions from inside the $Y_2(CO_3)_3 \cdot 2H_2O$ particles into the solution will happen. In contrast, the solutes ($CO_3^{2-}$, $Na^+$, $NH_4^+$) from the solution are more likely to stay at the entrance sites of the nanochannels (**Figures 6A, 6D, 6G**). Alternatively, the solutes can diffuse into the subsurface regions: e.g., a $\rightarrow$ a′ (rate constant $K_1'$ = 6.6 s$^{-1}$) for $CO_3^{2-}$; d $\rightarrow$ d′ ($K_2'$ = $1.7 \times 10^9$ s$^{-1}$) for $Na^+$; g $\rightarrow$ g′ ($K_3'$ = 0.74 s$^{-1}$) for $NH_4^+$ (**Figure 6**). It should be noted that the calculations have not taken into account the interactions between $CO_3^{2-}$ and $Na^+$ or $NH_4^+$. The attraction between the counter ions could help to lower the barriers and consequently facilitate subsurface diffusion of the solutes. The accumulation of net charges on the surfaces of the suspended particles will induce another driving force for the crystallization, as shown by previous work at the $H_2O$/NaCl system.[24] In our case, the adsorption and accumulation of $Na^+$ or $NH_4^+$ will attract $CO_3^{2-}$ to diffuse to the surfaces of $Y_2(CO_3)_3 \cdot 2H_2O$ particles and vice versa. For configurations near the surface of $Y_2(CO_3)_3 \cdot 2H_2O$ (**Figure S14**), we have calculated the strength of electrostatic attraction of $CO_3^{2-}$ — $Na^+$ to be ~ 1.96 eV, and $CO_3^{2-}$ — $NH_4^+$ to be ~ 2.27 eV. On the contrary, there is no such positive feedback effects in the diffusion of $Y^{3+}$ from inside $Y_2(CO_3)_3 \cdot 2H_2O$ to the surface, due to the screening effect of the coordination $CO_3^{2-}$ groups around each $Y^{3+}$.

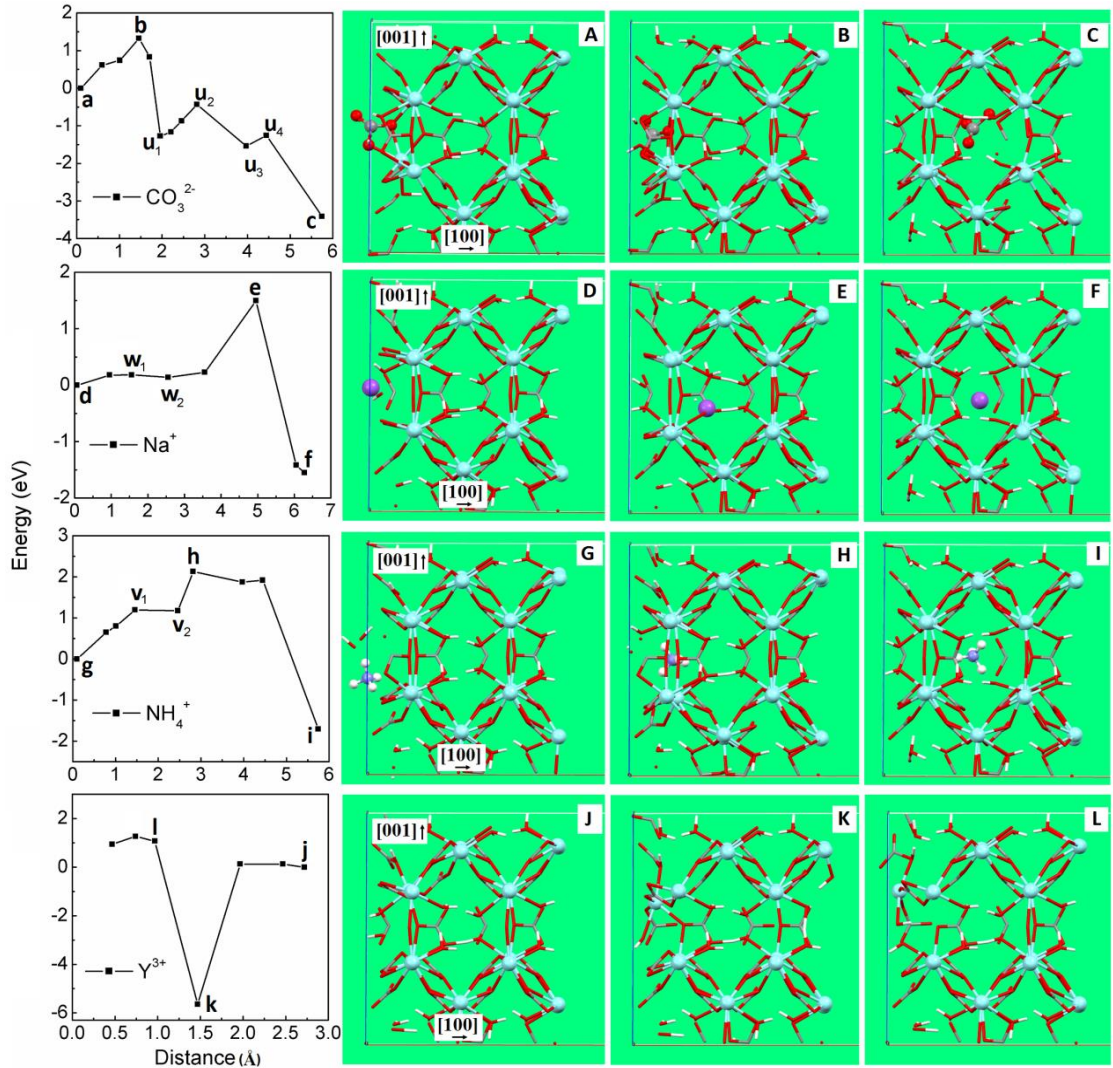

**Figure 6. Energetics and atomic configurations regarding the diffusion of ions. Left panels:** Relative energies of the atomic configurations as a function of the distance to the (100) surface of $Y_2(CO_3)_3 \cdot 2H_2O$. The configurations are located on the diffusion pathway for the ions from surface to the interior ($CO_3^{2-}$, $Na^+$, $NH_4^+$), or from the interior to surface ($Y^{3+}$), with the direction of diffusion schematically indicated by dashed arrows. For $CO_3^{2-}$ and $NH_4^+$, the distance is measured from the geometry center of the group to the surface. **Right panels:** Atomistic structures for some typical configurations, where the labeling capital letters **A-L** have a one-to-one correspondence to the energy configurations marked by the lowercase letters **a-l** in the left panels. For clear illustration, the $Y^{3+}$ ions (sky blue) and the ions/groups (indicated by vertical arrows) are represented by small balls (gray for C, purple for Na and violet for N), while other ions are represented by capped sticks in different colors: red for O and white for H.

Generally the modes of crystal growth can be classified into *two types*: the classical mode (monomer-by-monomer addition) and the particle attachment mode;[25] and OA belongs to the latter. The analysis above naturally points to the conclusion that it is the particle attachment instead of the classical mode of growth (dissolution-nucleation) that plays a dominant role in the formation of yttrium double carbonates. Variation of the frequency factor $A$ by one order of magnitude will not change the conclusion. Supposed that the contribution of classical mode of growth dominates, then the least time required for the formation of yttrium double carbonates is in the order of $10^{104}$ s ($\sim 10^{99}$ days), which is well beyond the time scale of the present experiment (1 $\sim$ 2 hours).

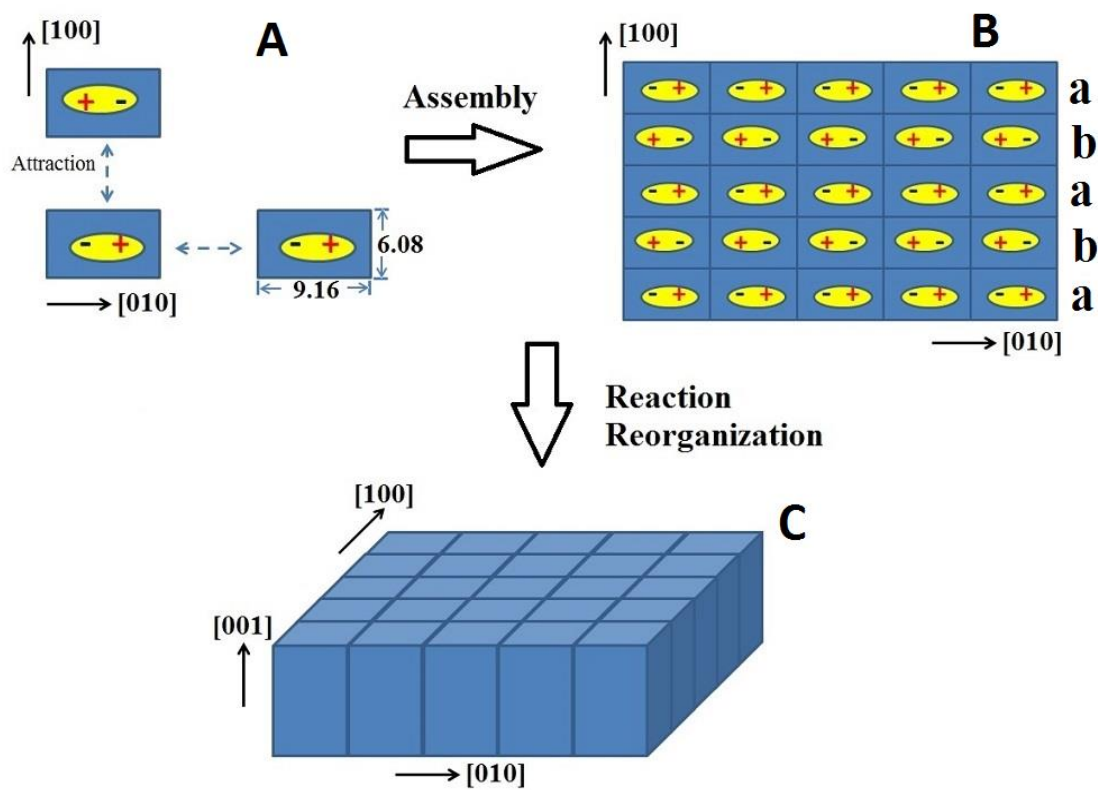

**Figure 7. Schematic diagram for the chemical reaction directed OA in yttrium double carbonates.** (*A*) Electrostatic interactions viewed from [001] direction. (*B*) Oriented assembly of the nanoparticles via electrostatic interactions. (*C*) Formation of the 2D-like sheets by reaction and position reorganization of the ions.

It is fascinating how the suspended particles with adsorbed ions are assembled to form the single-crystalline, 2D-like sheets obtained in our experiments. Besides the stresses associated with lattice mismatch, which tend to induce rotation of the nanoparticles to contact at surfaces with the same orientation and to form aligned configurations, the most important driving force is the electrostatic interactions between nanoparticles.[9,26] As mentioned above, to maintain charge neutrality, the ions which firstly adsorbed at the entrance sites of the nanochannels will attract the ions with opposite charges from the solution to the surfaces of the nanoparticles. Consequently, the two neighboring nanochannels will contain net charges with opposite signs ($\pm 1$). This forms electric dipoles. **Figure 7** schematically shows the OA process of the nanoparticles via dipole-diploe interactions. The smallest distances ($L_m$) between the dipoles are limited by the dimensions of the unit cell of $Y_2(CO_3)_3 \cdot 2H_2O$, which is $L_m \sim 6.08$ Å, 4.58 Å (one half of the b-axis of unit cell), 15.11 Å in the X, Y, Z direction, respectively.[17] To account for the screening effects from the other solutes in the solution, we have estimated the Debye length ($\lambda_D$) at the initial stage of OA growth, which is $\sim 5.09$ Å (**Supporting Information** Text, Theoretical Part C). At distances beyond $\lambda_D$ the electrostatic interactions decay exponentially and Brownian motions will be dominant. This is in line with our experimental observation that in a dilute solution the reactions will cease, due to the very low collision probability of the nanoparticles. From the values of $L_m$ and $\lambda_D$, the electrostatic interactions in the Z direction will be much weaker than in the X & Y directions. Therefore, the assembly of the nanoparticles will proceeds mainly along X and Y directions while the packing along the Z direction will be much slower. This picture holds for the alignment of nanoparticles which contain a number of unit cells, since the dominant dipole-dipole interactions along each direction are still determined by the lattice constants. As shown in **Figure 7**, the Coulomb interactions between the nanoparticles will result in an "abab" stacking order in the X ([100]) direction, where the "b" lines are just rotated by 180° from "a" lines. Reaction and reorganization (details at atomic level are left for future studies) of the ions in the assembled units will lead to the fusion of the interfaces of crystallites and get the final products: the 2D-like sheets.

**CONCLUSIONS**

In summary, by combining experimental and theoretical efforts, we demonstrate that with the participation of $NaHCO_3$ and $NH_4HCO_3$, micron-sized sheets of $NaY(CO_3)_2 \cdot 6H_2O$ and $(NH_4)Y(CO_3)_2 \cdot H_2O$ can be formed from $Y_2(CO_3)_3 \cdot 2H_2O$ nanoparticles through a chemical reaction directed oriented attachment. In contrast to conventional OA process, this OA growth can produce new single-crystalline substances whose crystallographic structures and chemical constituents are different (usually more complicated) from that of the precursor nanoparticles. Such an OA mechanism shows applicability in a number of salts. Our findings not only extend the territory of the nonclassical aggregation-based crystal growth mode, but also provide a feasible approach for creating new functional materials with complex architectures from simple nanoparticles.

**METHODS**

1) **Nanoparticles synthesis and oriented attachment.** Several groups of $Y_2(CO_3)_3 \cdot 2H_2O$ nanoparticle suspensions (labeled as Y0) were synthesized first by dripping aqueous $NH_4HCO_3$ solution (1.0 M, 10 ml) slowly into the mixed solution (20 ml, pre-heating at 60 °C) of $Y(NO_3)_3$ (0.2 M) and $(NH_4)_2SO_4$ (0.01 M) with stirring. Afterwards, aqueous solutions (1.0 M, 15 ml) of $NaHCO_3$, $NH_4HCO_3$, $NaOH$, and $NH_4OH$ were dripped into the above $Y_2(CO_3)_3 \cdot 2H_2O$ suspension at a rate of 1 ml/min under vigorous agitation, respectively. After reaction for 10-120 min, the white products (labeled as Y1, Y2, Y3, and Y4, respectively) were harvested and washed thoroughly with deionized water and absolute ethyl alcohol successively. The similar procedure was also used for the preparation of $NaGd(CO_3)_2 \cdot 6H_2O$ and $KNd(CO_3)_2 \cdot xH_2O$ micron-sized sheets.

2) **Post-treatment to the micron-sized sheets.** Typically, $(NH_4)Y(CO_3)_2 \cdot H_2O$ suspensions (3 ml) were mixed with 3 ml $NH_4OH$ solution (1.0 M) into 45 ml of solvents (ionized water, absolute ethyl alcohol, and toluene, respectively). The obtained mixtures were then aged at 160 ℃ for 24 hrs in a Teflon autoclave (65 ml) respectively.

3) **Characterization.** The particle size distribution of $Y_2(CO_3)_3 \cdot 2H_2O$ nanoparticles was measured by small angle X-ray scattering (SAXS, Philips X'Pert Pro MPD). The phase, microstructure, and morphology of the products were investigated using X-ray diffraction (XRD, Philips X'pert, $Cu_{k\alpha}$), field-emission scanning electron microscopy (FESEM, Sirion 200), transmission electron microscopy (TEM, JEM-2010), and tapping-mode atomic force microscopy (AFM, Seiko, SPA-300HV & SPI3800N), respectively. The yttrium content of $(NH_4)Y(CO_3)_2 \cdot H_2O$ micron-sized sheets was analyzed by inductively coupled plasma emission spectrometer (ICP, Thermal Electron). Carbon, Hydrogen, and Nitrogen contents of $(NH_4)Y(CO_3)_2 \cdot H_2O$ micron-sized sheets were determined by an elemental analyzer (Vario EL cube, Elementar). Fourier transform infrared (FTIR) spectroscopy of the samples was accomplished using a Nexus Nicolet spectrometer. The EXAFS measurements for the Y0 suspension, Y2 suspension, and the suspension sampled during the formation process of Y2, were performed at the beamline 14W1 of the Shanghai Synchrotron Radiation Facility (SSRF). The X-ray was monochromatized by a double-crystal Si (311) monochromator for SSRF. The monochromator was detuned to reject higher harmonics. All suspensions were in-situ measured in the transmission mode.

4) **First-principles calculations.** Our first-principles calculations are carried out by the Vienna *ab initio* simulation package (VASP),[27, 28] which is based on density functional theory (DFT). A plane wave basis set and the projector-augmented-wave (PAW) potentials[29, 30] are employed to describe the electron-ion interactions. The exchange-correlation interactions of electrons are described by the PBE functional.[31] The energy cutoff for plane waves is 600 eV. The primitive cell of crystalline $Y_2(CO_3)_3 \cdot 2H_2O$ is constructed based on the experimental work,[17] which consists of 4 formula units of $Y_2(CO_3)_3 \cdot 2H_2O$. To simulate the diffusion of ions (from surface to interior ($CO_3^{2-}$, $Na^+$, $NH_4^+$) or interior to surface ($Y^{3+}$)), we take a ($2 \times 1 \times 1$) unit cell (totally 160 ions) to model the $Y_2(CO_3)_3 \cdot 2H_2O$ nanoparticles, with the (100) surface separated by a vacuum layer of ~ 15 Å (the conclusion drawn in this paper is maintained for calculations using a vacuum layer of 20 Å (**Figure S15**). The diffusion of ions takes place along the [100] direction, the direction of the nanochannel in $Y_2(CO_3)_3 \cdot 2H_2O$. A $1 \times 2 \times 1$ k-mesh generated using the Monkhorst-Pack scheme is employed for all the calculations.[32] For each ionic configuration along the diffusion pathway, the x coordinates of the geometric center of the diffusion ion or ion clusters are fixed with their y & z coordinates relaxed. To simulate the bulk environment and to remove the artificial perturbations along the [100] direction due to periodic boundary conditions, similar restriction is applied to the relaxation of the ions which locate outside the first coordination shell of the diffusion ions. To verify the reliability of the obtained energy barriers, we have also studied the diffusion of $Y^{3+}$ using the Nudged Elastic Band (NEB) method,[33,34] which turns out to be a huge computational burden for DFT simulations. The computed energy barrier (~ 7.15 eV) of the rate-determined step along the diffusion pathway is shown in Figure S16, which is very close to the value (~ 6.91 eV) obtained using the step-by-step relaxation method (**Figure 6**). The aqueous environment surrounding the $Y_2(CO_3)_3 \cdot 2H_2O$ nanoparticles is not considered in our simulation due to the following two reasons: i) The diffusion of ions in liquids is much faster/easier than in solids; and ii) The inclusion of liquid water is too much demanding for the present first-principles simulations.

**ACKNOWLEDGMENTS**

This work was supported in part by National Natural Science Foundation of China (No. 11374306, No. 11174292, No. 11674322, and No. 51672278). The authors thank Professors Lide Zhang and Guowen Meng for helpful discussions; Professor Xiaojia Chen for reading of the manuscript; Li Chen and Zhaoqin Chu for their help with SAXS and TEM analyses. Y. F. Liu is grateful to Dr. Junming Xu, Zhen Gu and Hualing Ding for helpful reference and discussion. The DFT simulations are performed using the supercomputers of the Hefei Branch of Supercomputing Center of Chinese Academy of Sciences; the supercomputers of the Institute for Materials Research, Tohoku University; and the Tian He II series supercomputers. Y.Y. acknowledges support from the National Natural Science Foundation of China (No. 11474285).

# ASSOCIATED CONTENT

# AUTHOR INFORMATION

Correspondence and requests for materials should be addressed to X.Y.Q (xyqin@issp.ac.cn), Y.F.L. (liuyongfei.mse@gmail.com), and Y.Y. (yyang@theory.issp.ac.cn; yyangtaoism@gmail.com).

**Notes**

The authors declare no competing financial interests.

**Supporting  Information**

**Contents:**

SI Figures S1 to S19
SI Table S1
SI Text
SI References (35-48)

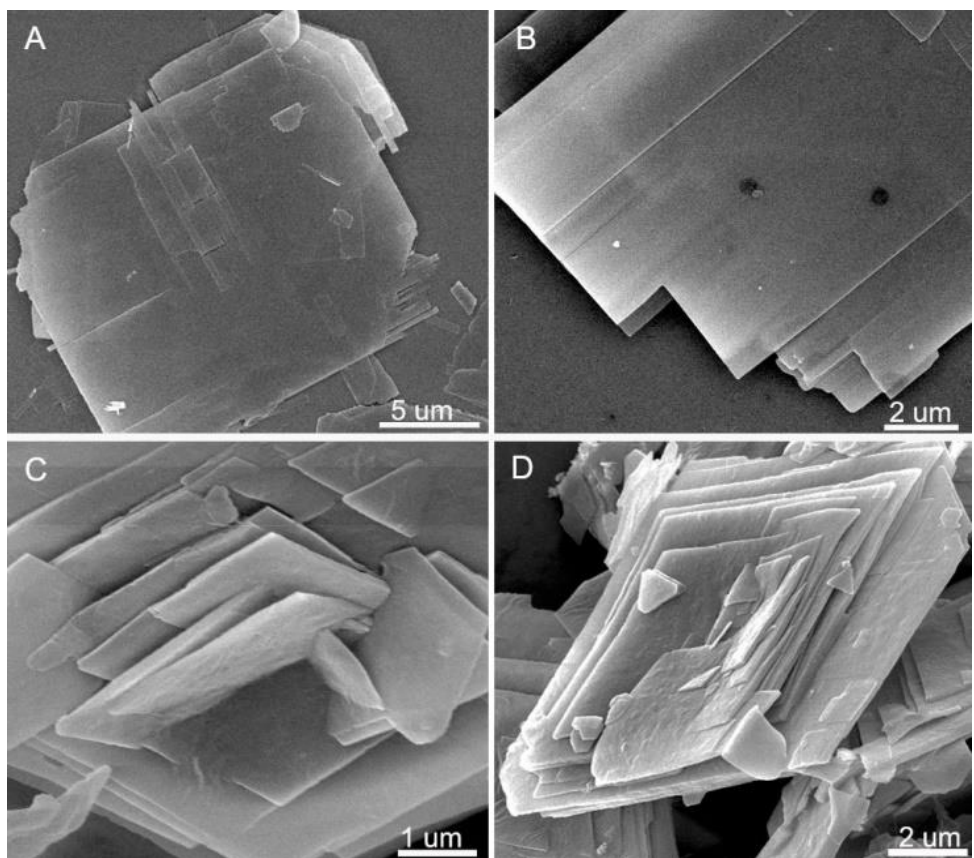

**Figure S1.** FESEM images of the obtained sheets. (*A-B*) NaY(CO$_3$)$_2$·6H$_2$O, (*C-D*) (NH$_4$)Y(CO$_3$)$_2$·H$_2$O.

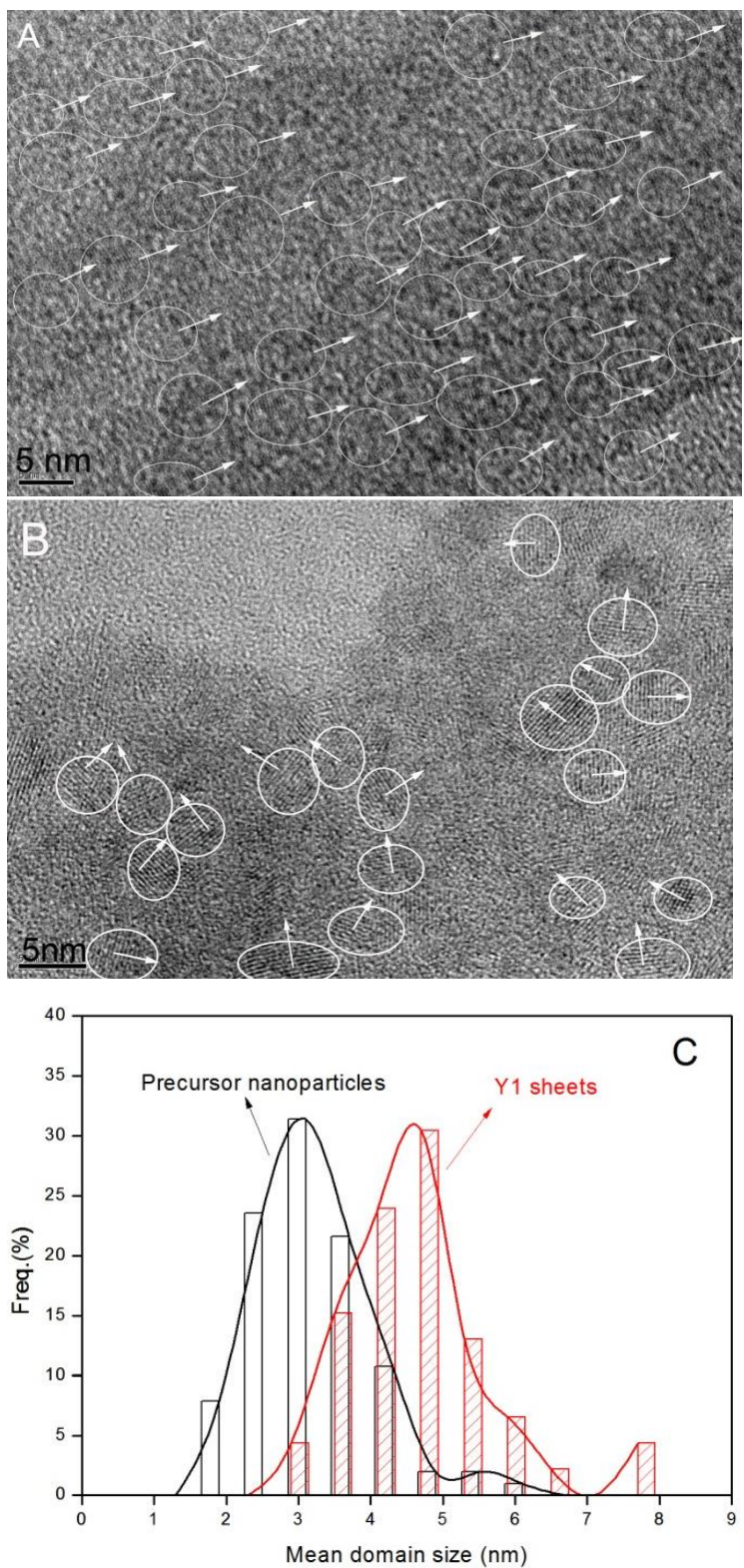

**Figure S2.** High resolution TEM images of the NaY(CO$_3$)$_2$·6H$_2$O sheets (*A*) and precursor Y$_2$(CO$_3$)$_3$·2H$_2$O particles (*B*); The domain size distributions of the precursor nanoparticles and Y1 sheets (*C*).

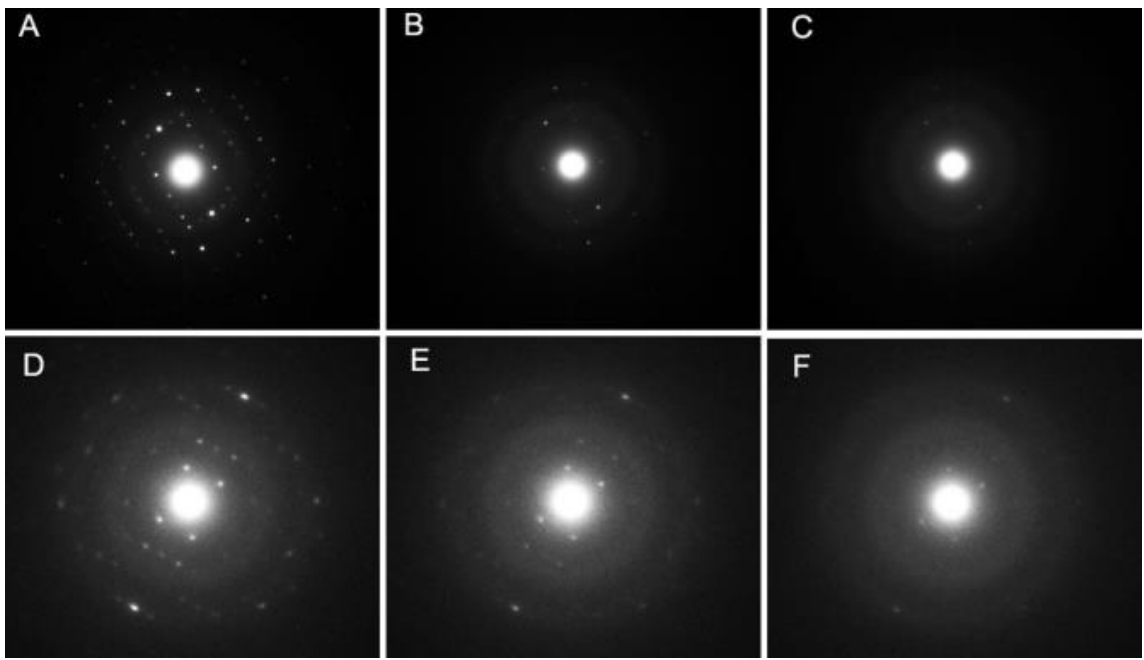

**Figure S3.** Evolution of SAED patterns of (*A-C*) NaY(CO$_3$)$_2$·6H$_2$O sheet within ~ 30 seconds and (D-F) (NH$_4$)Y(CO$_3$)$_2$·H$_2$O sheet within ~ 10 seconds due to the beam damage. Such rapid decomposition of (NH$_4$)Y(CO$_3$)$_2$·H$_2$O sheet leads to difficulty to obtain the lattice-fringe image.

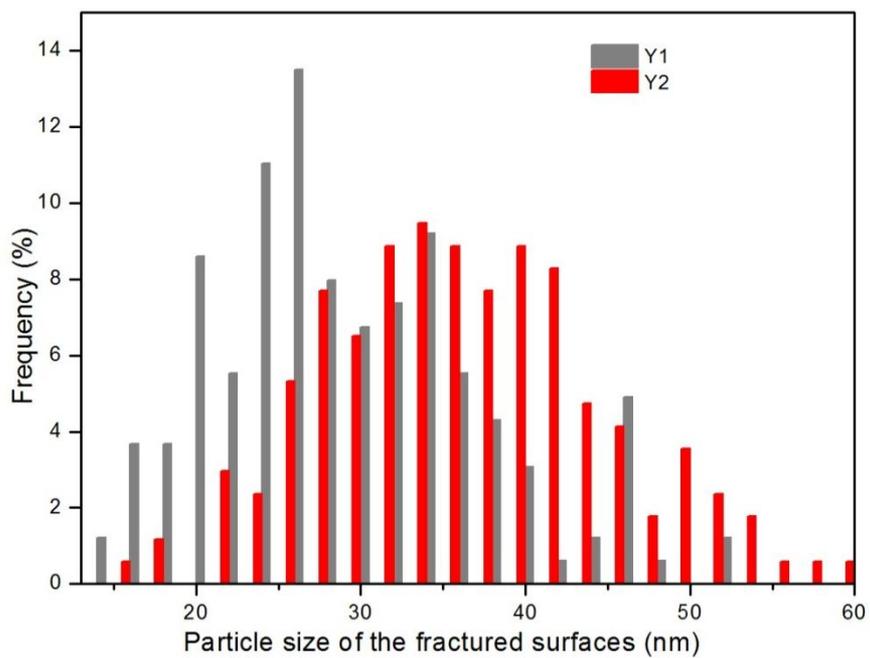

**Figure S4.** Particle size distributions on the fracture surfaces of Y1 and Y2.sheets, as measured from AFM images.

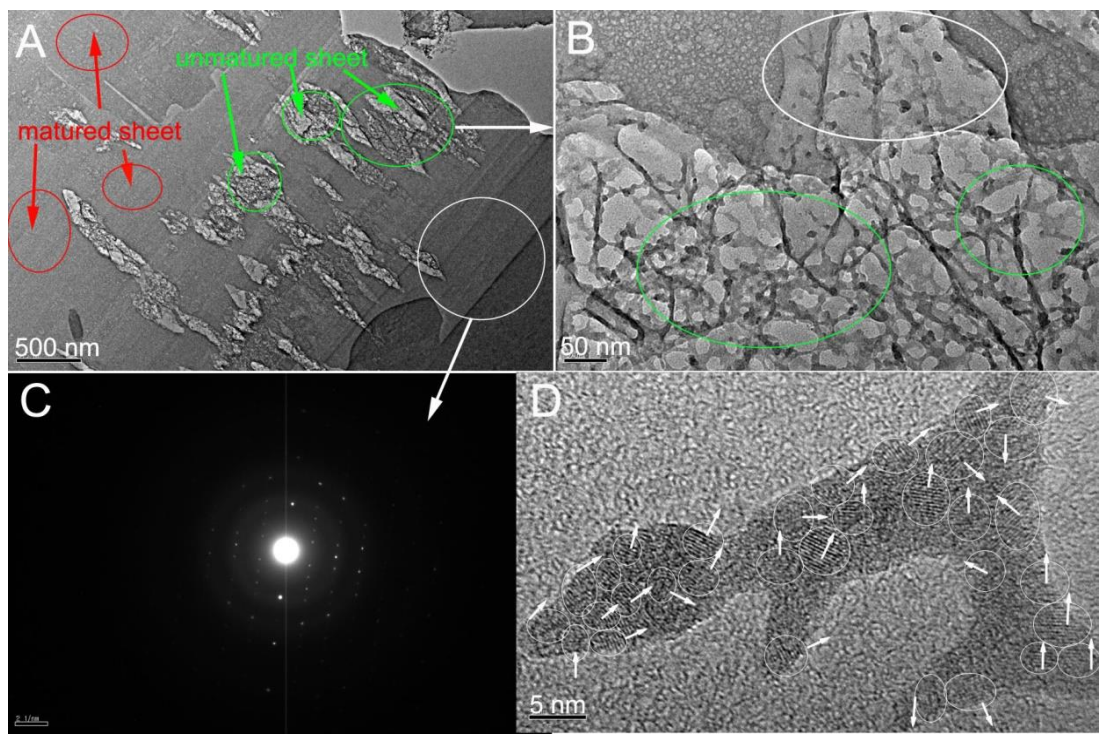

**Figure S5.** TEM images (*A* and *B*), SAED pattern (*C*), and HRTEM image (*D*) for the intermediates in the formation of NaY(CO$_3$)$_2$·6H$_2$O sheets.

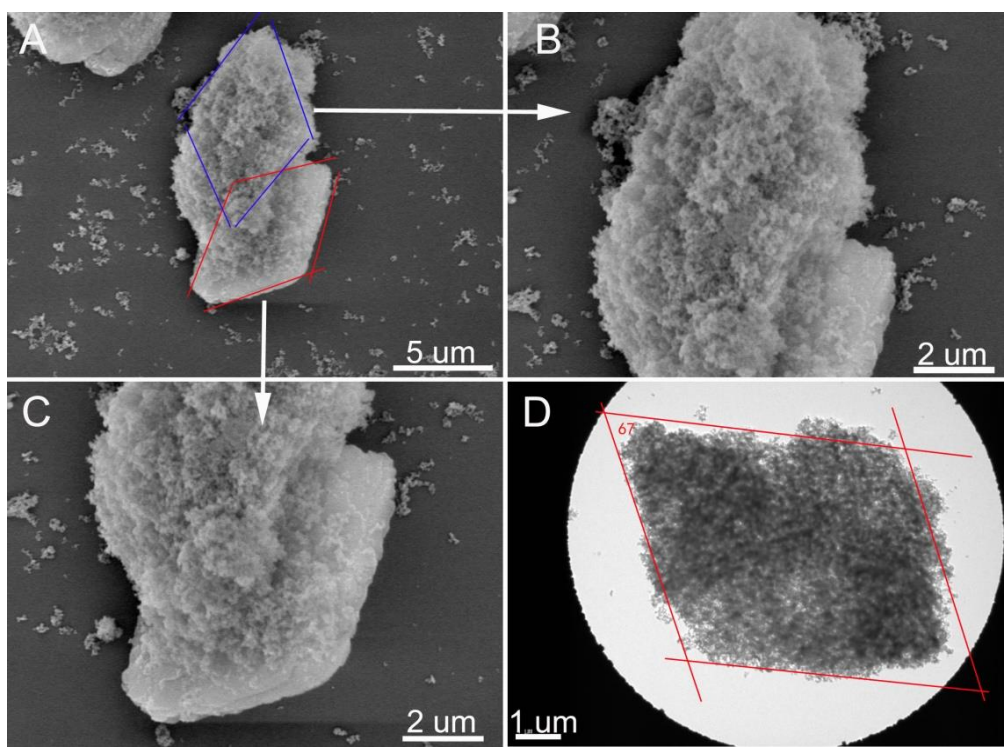

**Figure S6.** FESEM (*A-C*) and TEM (*D*) images for the intermediates in the formation of (NH$_4$)Y(CO$_3$)$_2$·H$_2$O sheets.

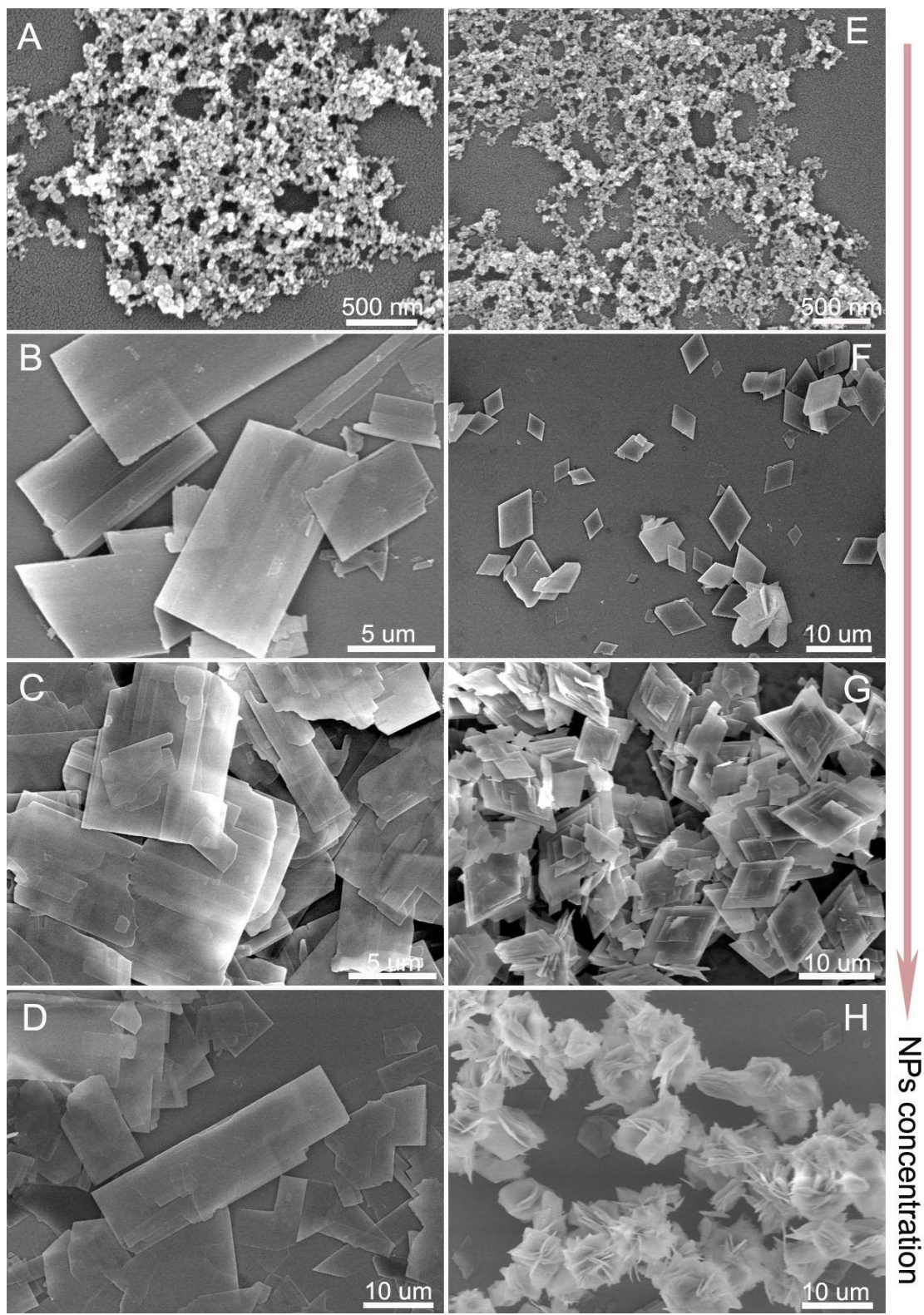

**Figure S7.** Morphologies of Y1 (left panels) and Y2 (right panels) evolving with the NPs concentration. FESEM images of the products synthesized in the condition of diluting Y0 suspensions, NaHCO$_3$ and NH$_4$HCO$_3$ solutions synchronously to 1/50 (*A* and *E*), 1/10 (*B* and *F*), 1/1 (*C* and *G*), and a condensed Y0 suspension (*D* and *H*, Y0 was synthesized from 0.5M Y(NO$_3$)$_3$ solution) with a reaction time for 48 hrs, respectively.

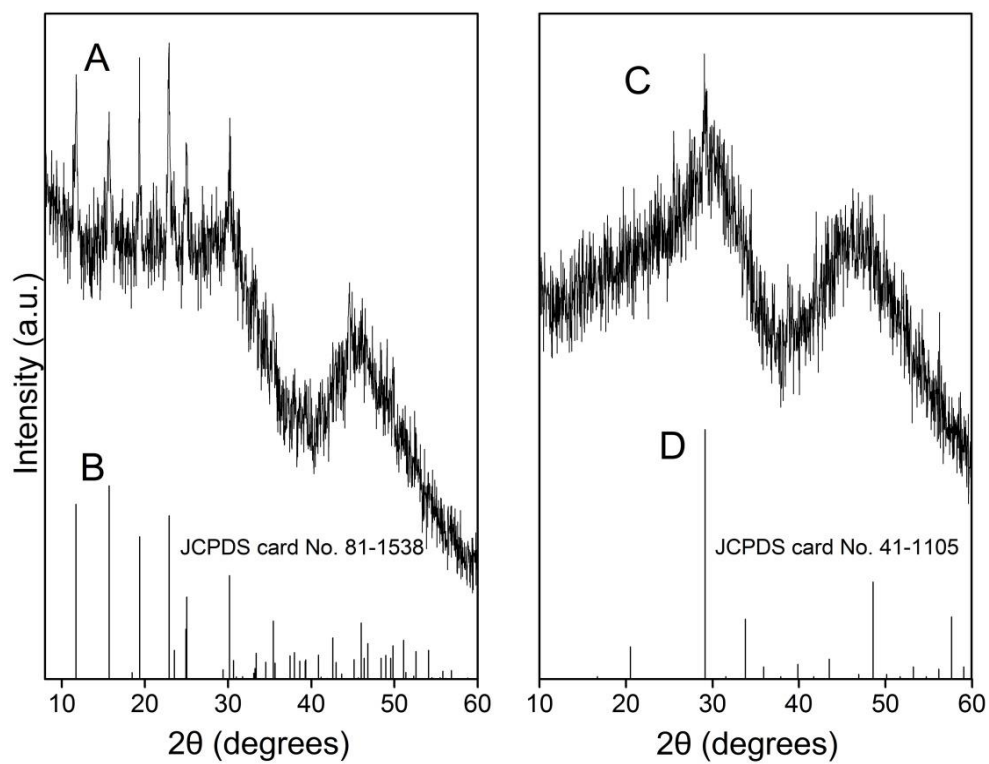

**Figure S8.** XRD patterns of the $(NH_4) \cdot Y(CO_3)_2 \cdot H_2O$ sheets after solvothermal treatment in ethanol at 160 °C for 4 hrs (*A*) and 24 hrs (*C*). Therein (*B*) is orthorhombic $Y_2(CO_3)_3 \cdot 2H_2O$ pattern (JCPDS No. 81-1538) and (*D*) is cubic $Y_2O_3$ pattern (JCPDS No. 41-1105).

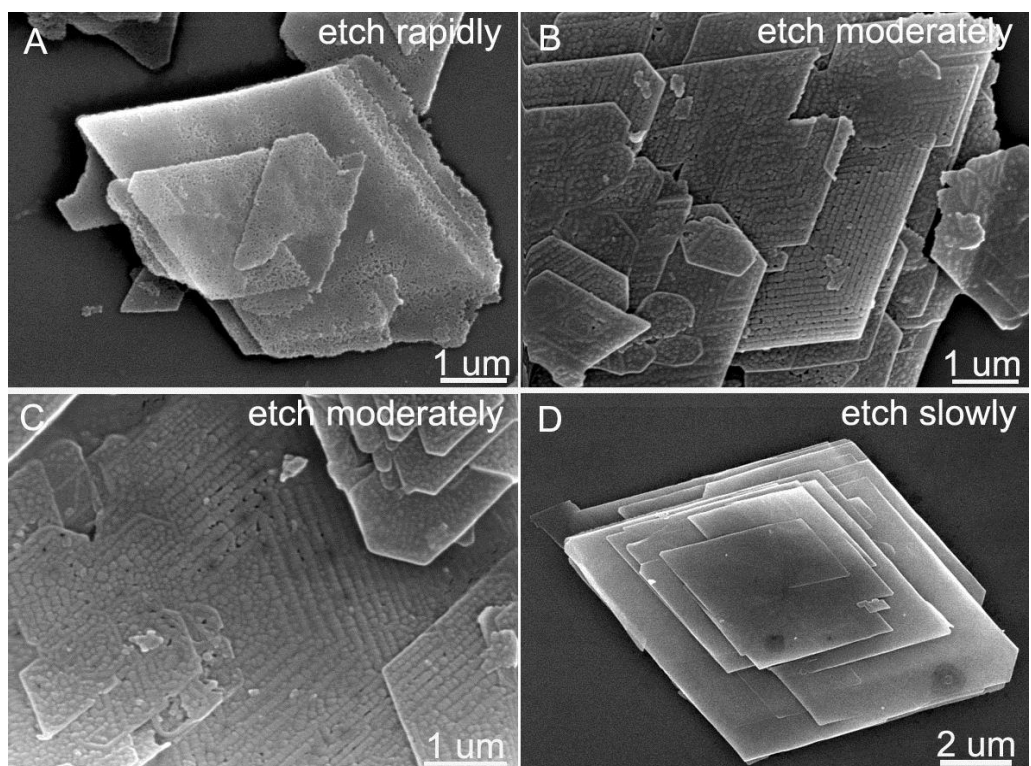

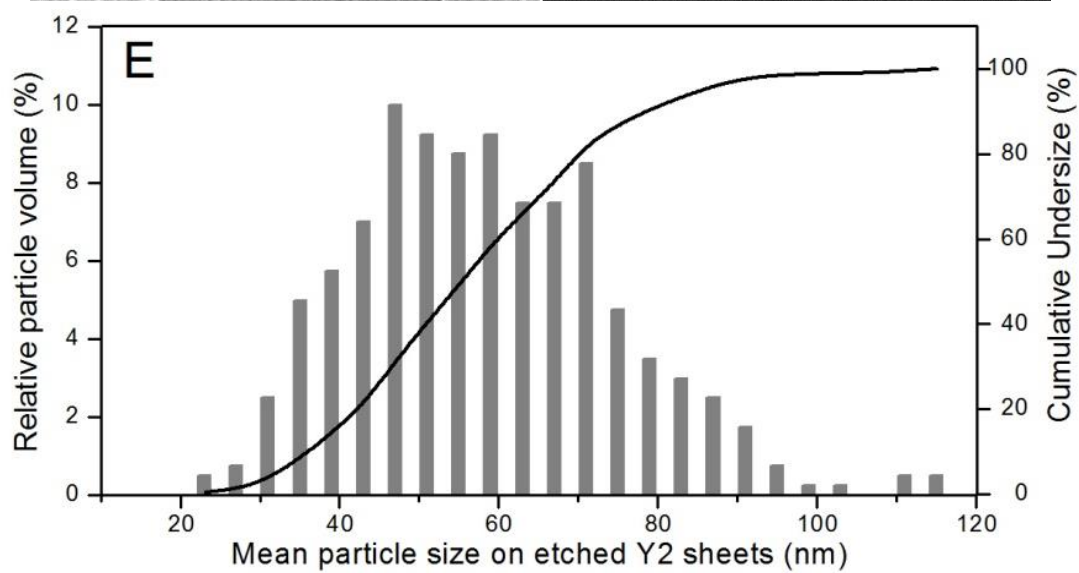

**Figure S9.** FESEM images of the $(NH_4)\cdot Y(CO_3)_2\cdot H_2O$ sheets after solvothermal treatment at 160 °C for 24 hrs in ethanol with different addition of $NH_3\cdot H_2O$: 0 mL (*A*), 3 mL (*B* and *C*), and 6 mL (*D*). (*E*) The size distribution of particles in the etched Y2 sheets obtained at the moderate treatment conditions.

As the XRD examination revealed (**Figure S7**), during the solvothermal etching treatment of the $(NH_4)\cdot Y(CO_3)_2\cdot H_2O$ sheets, following chemical reactions will occur:

$$2(NH_4)\cdot Y(CO_3)_2\cdot H_2O \rightarrow Y_2(CO_3)_3\cdot 2H_2O + 2NH_4^+ + CO_3^{2-} \quad (5)$$

$$Y_2(CO_3)_3\cdot 2H_2O \rightarrow Y_2O_3 + 3CO_2 + 2H_2O \quad (6)$$

Theoretically, the reaction will take place easily in the defect-rich interface regions. In order to make the defect-rich regions show up clearly, a moderate reaction is required to insure that reaction (5) occurs at these regions preferentially. The chemical reaction was tuned by altering the addition of $NH_3\cdot H_2O$ (supplying $NH_4^+$), which can decelerate the decomposition reaction by increasing the concentration of $NH_4^+$ to accelerate the reverse reaction of (5). As shown in **Figure S8**, with increasing addition of $NH_3 \cdot H_2O$, the dissolution rate slows down gradually, which allow us to obtain the products with disordered wormhole-like porous structure, the products with regular particle array patterns, and the products with smooth surfaces, respectively. The change trend in the surface morphologies of the etched sheets is in good agreement with our above analysis, showing the validity of the etching.

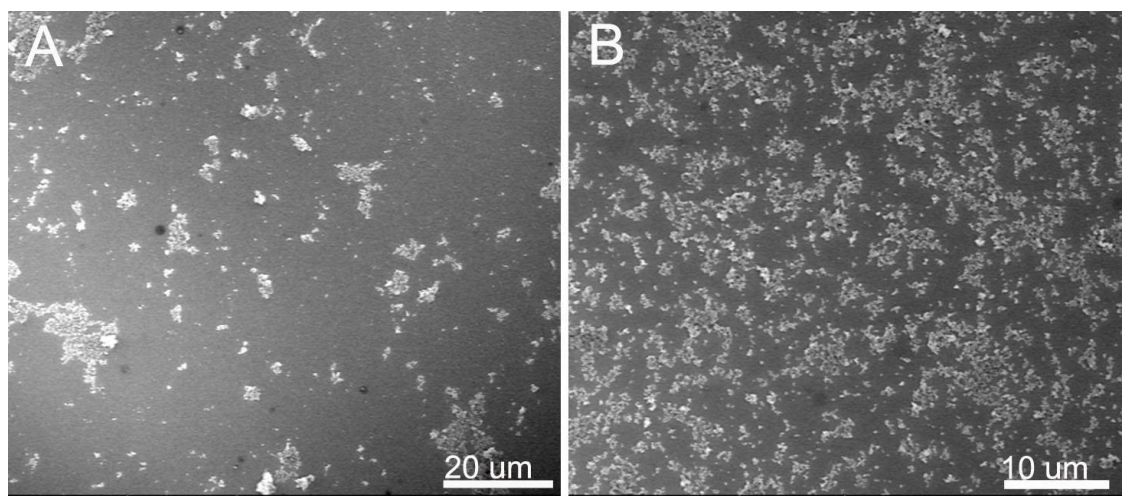

**Figure S10.** FESEM images of the samples synthesized by introducing (*A*) NaOH and (*B*) NH$_4$OH aqueous solutions into $Y_2(CO_3)_3 \cdot 2H_2O$ nanoparticle suspensions. No micron-sized or nano-sized sheets were obtained, indicating the key role of the double-salt reaction in the OA process.

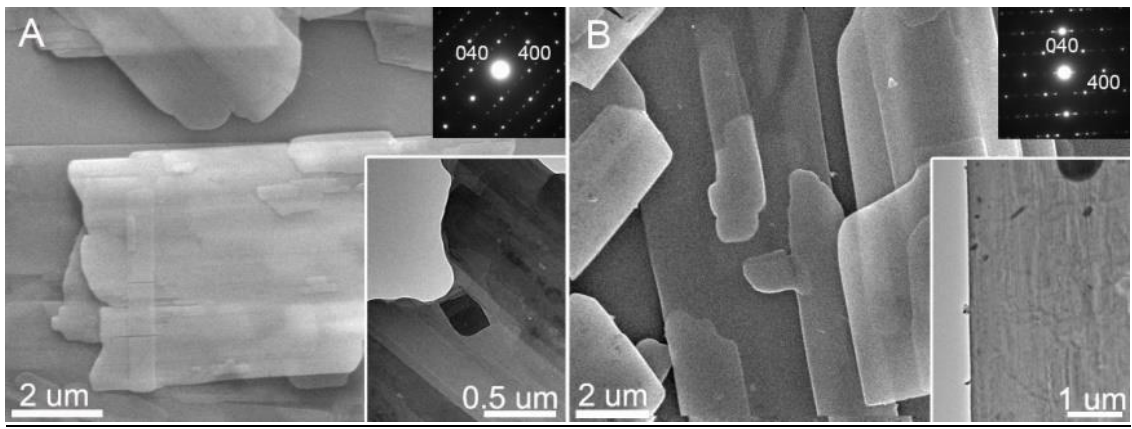

**Figure S11.** FESEM images of the micron-sized sheets synthesized via the OA process: (*A*) NaGd(CO$_3$)$_2$ 6H$_2$O (JCPDS card No. 31-1291); (*B*) KNd(CO$_3$)$_2$ xH$_2$O. Insets are corresponding TEM images and SEAD patterns. The SAED patterns show single-crystalline nature of the sheets, indicating that the formation of these multilayer sheets is not from the simple and random packing of single-layer sheets, but the result of well-organized layer growth through the OA of nanoparticles. The SAED pattern of KNd(CO$_3$)$_2$ xH$_2$O is indexed according to the data (Tetragonal, *a*=13.28 Å, *c*=10.00 Å) from ref [35].

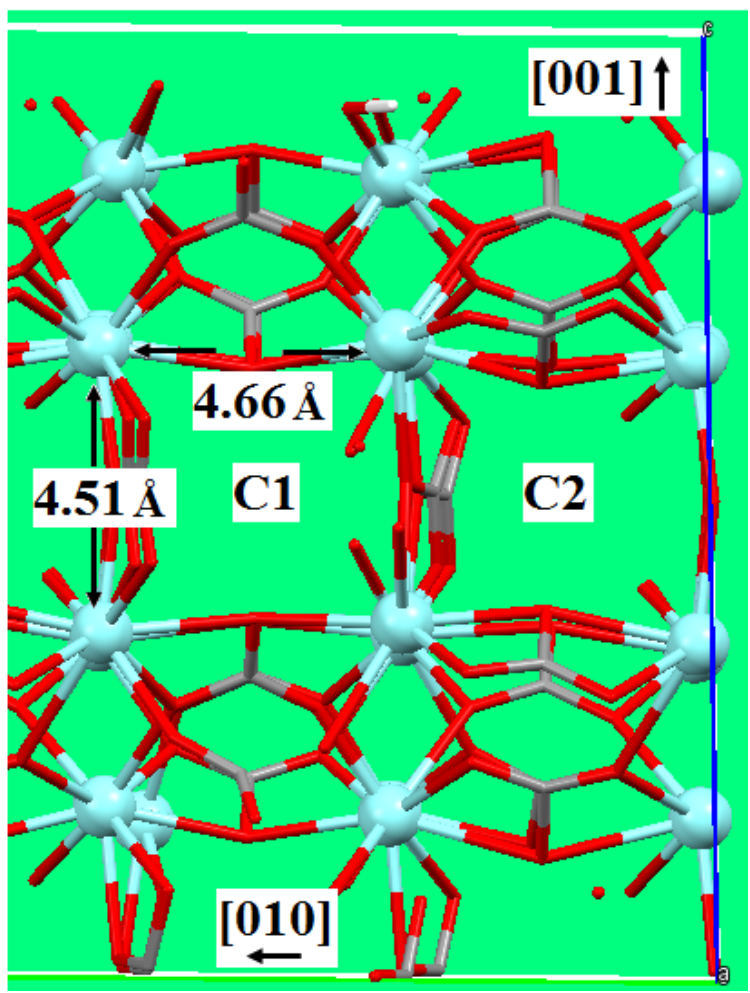

**Figure S12.** Atomistic structures of the nanochannels (C1 and C2) in $Y_2(CO_3)_3 \cdot 2H_2O$ for the diffusion of $CO_3^{2-}$, $Na^+$, and $NH_4^+$.

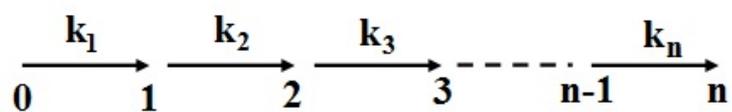

**Figure S13. Schematic diagram for a multistep diffusion process with the rate constant $k_i$ for step $i$.**

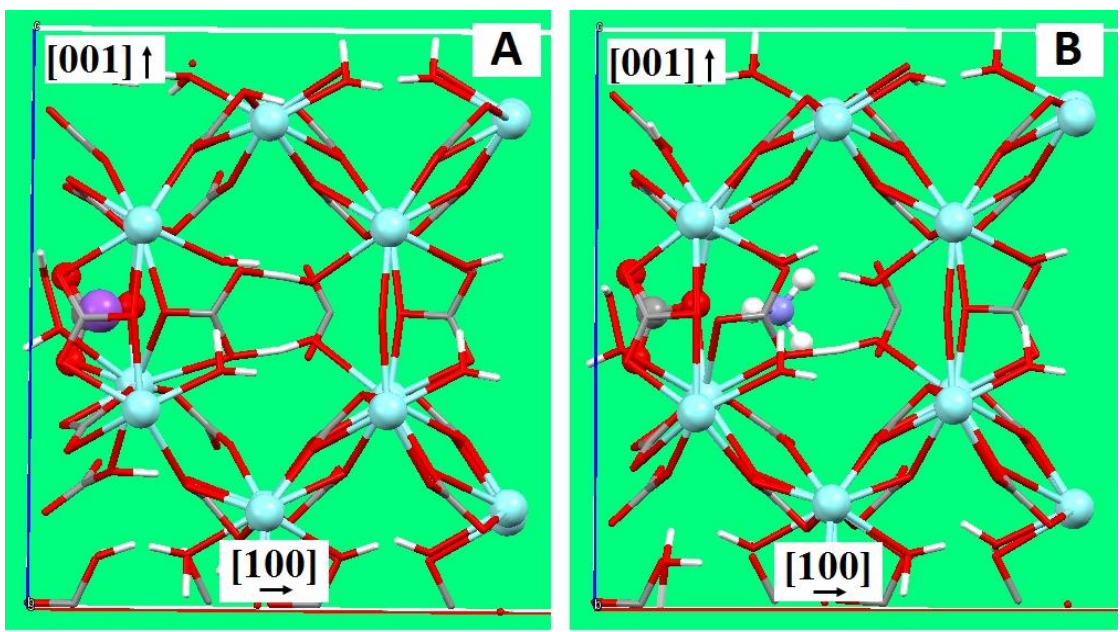

**Figure S14. Atomistic models for the calculation of electrostatic interactions.** (*A*). Interaction between $CO_3^{2-}$ and $Na^+$. (*B*). Interaction between $CO_3^{2-}$ and $NH_4^+$. $CO_3^{2-}$ locates near the surface region. The presentations of balls and capped sticks have the same meaning as in **Figure 6**.

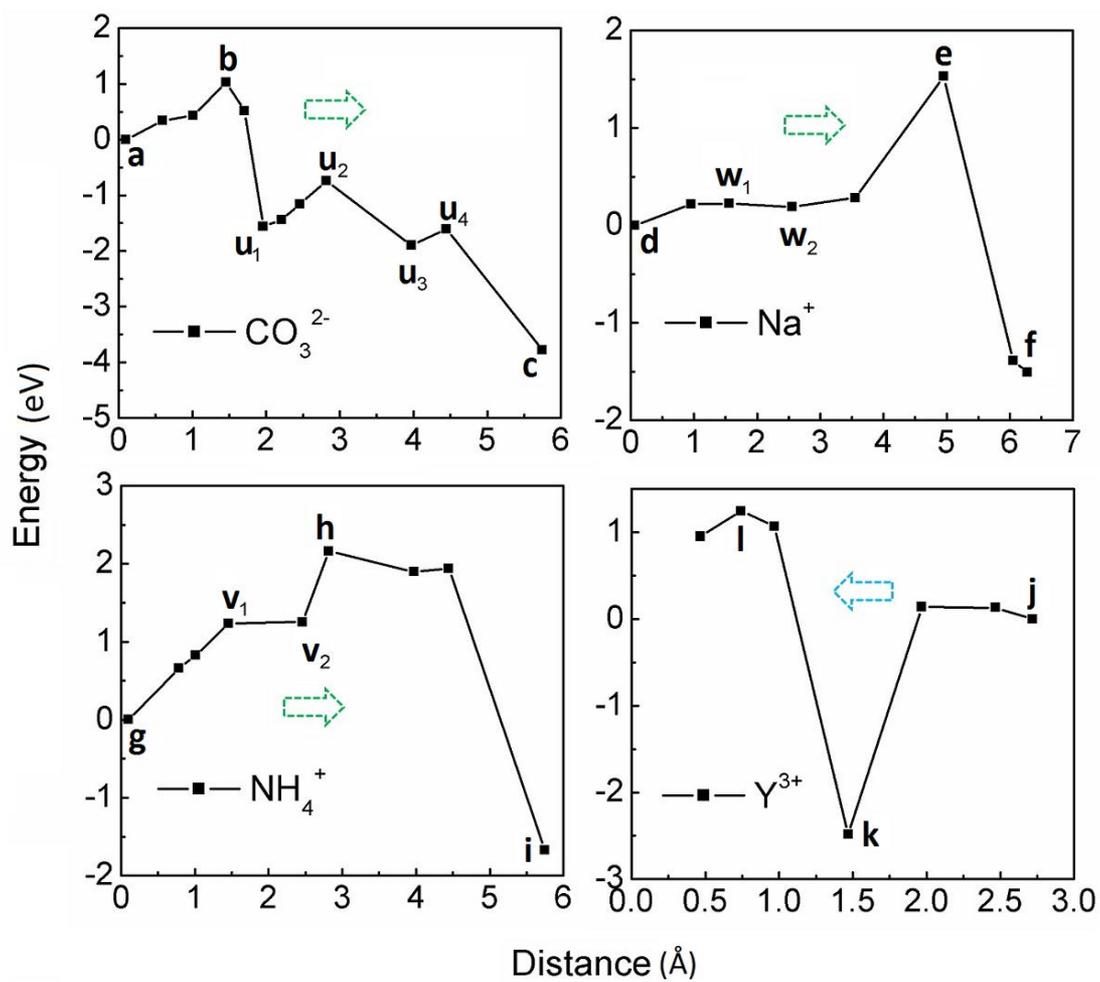

**Figure S15.** Energetics regarding the diffusion of the ions/ion groups, calculated using a vacuum layer of 20 Å for separating the simulation slab.

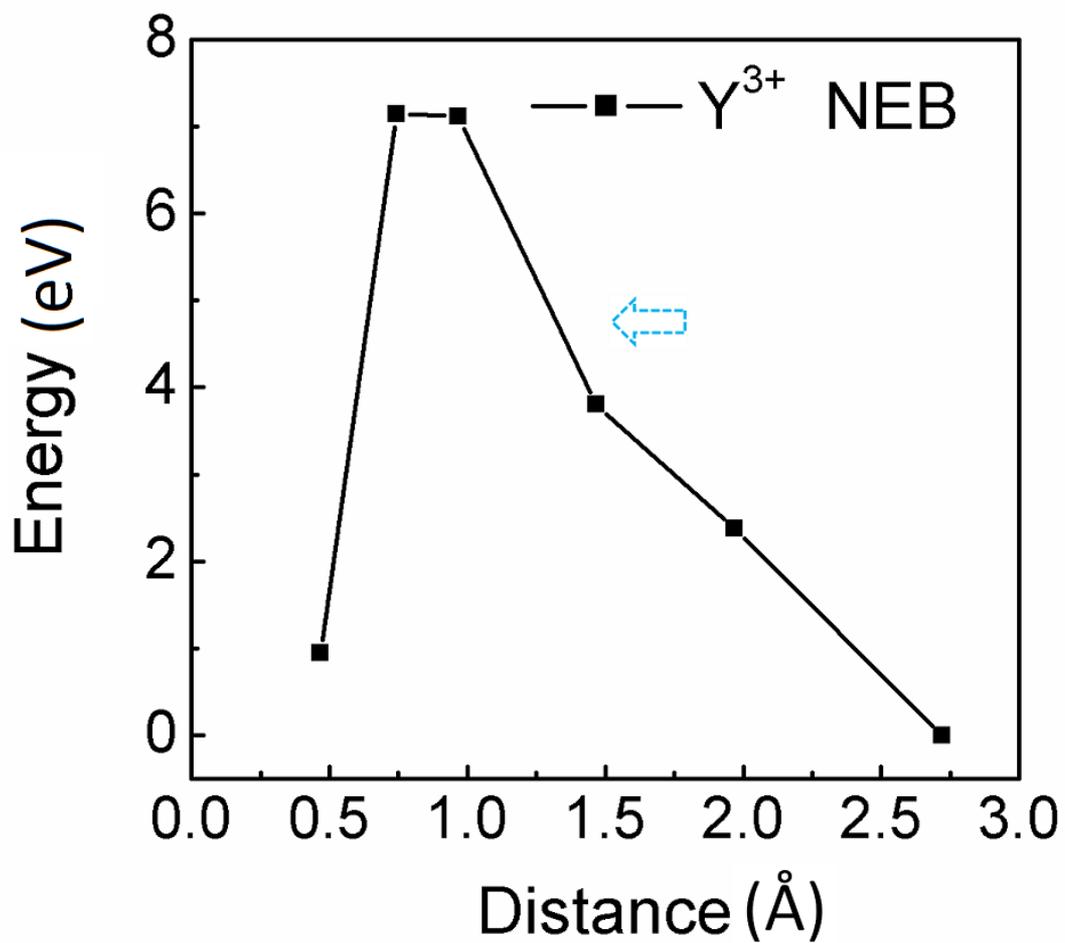

**Figure S16.** Energetics regarding the diffusion of the $Y^{3+}$, calculated using the Nudged Elastic Band (NEB) method, and a vacuum layer of 15 Å for separating the simulation slab.

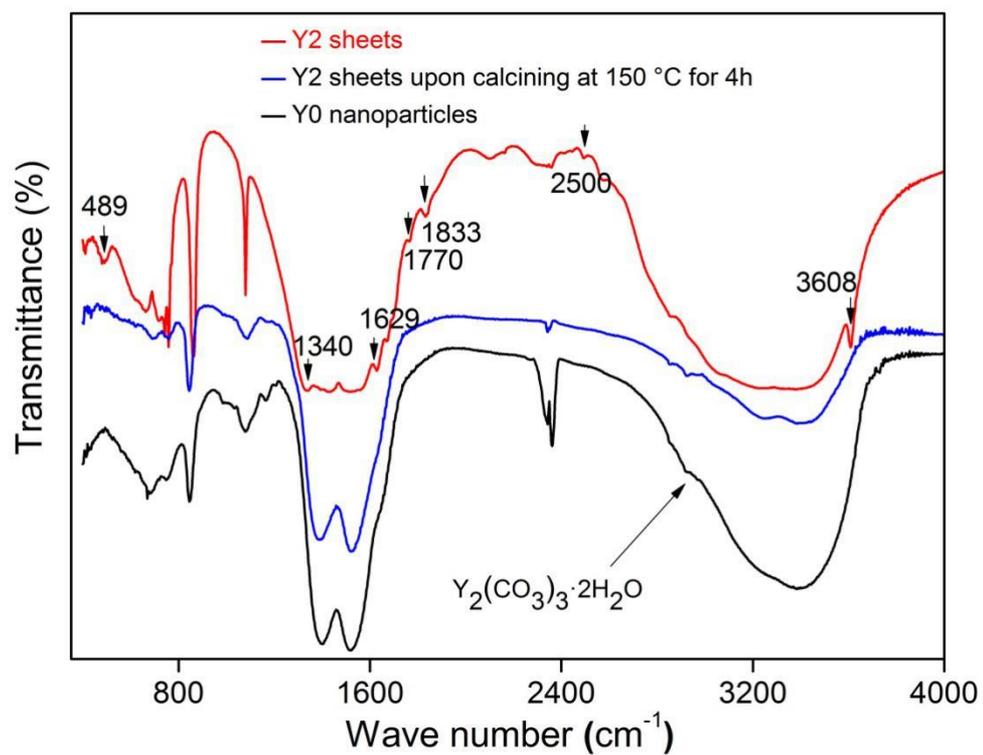

**Figure S17. FTIR spectra of the samples.** The peaks marked by the arrows verify the existence of $NH_4^+$ in Y2 ($(NH_4) \cdot Y(CO_3)_2 \cdot H_2O$) micron-sized sheet.

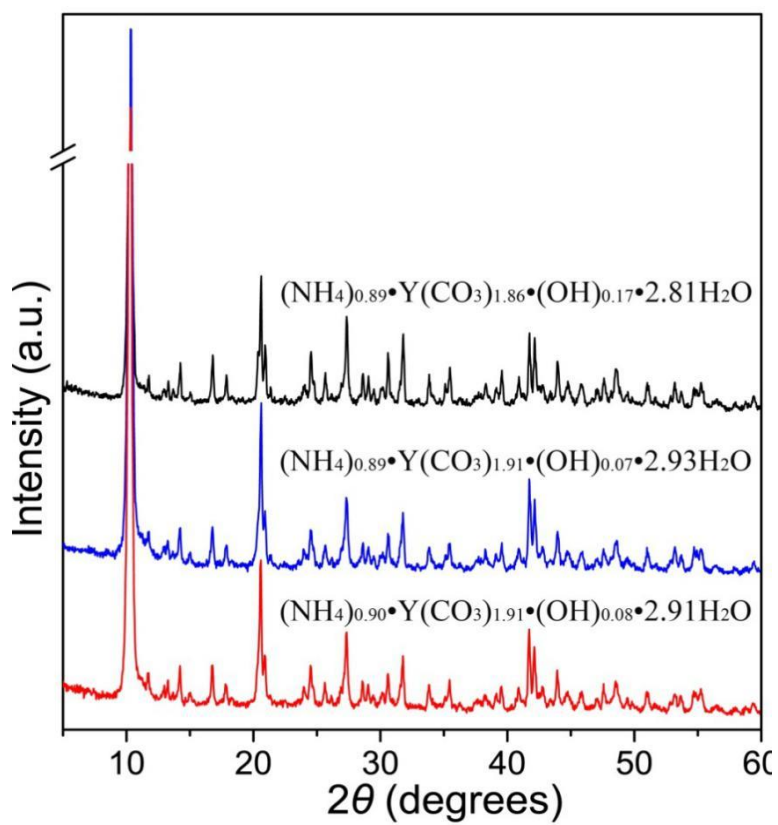

**Figure S18.** Powder XRD patterns of Y2 micron-sized sheets with slight differences in chemical compositions synthesized under different molar ratio of total $NH_4HCO_3$ to $Y(NO_3)_3$ (see **Table S1**). No obvious differences in the patterns are observed, which indicate that there is no apparent change in its crystal structure.

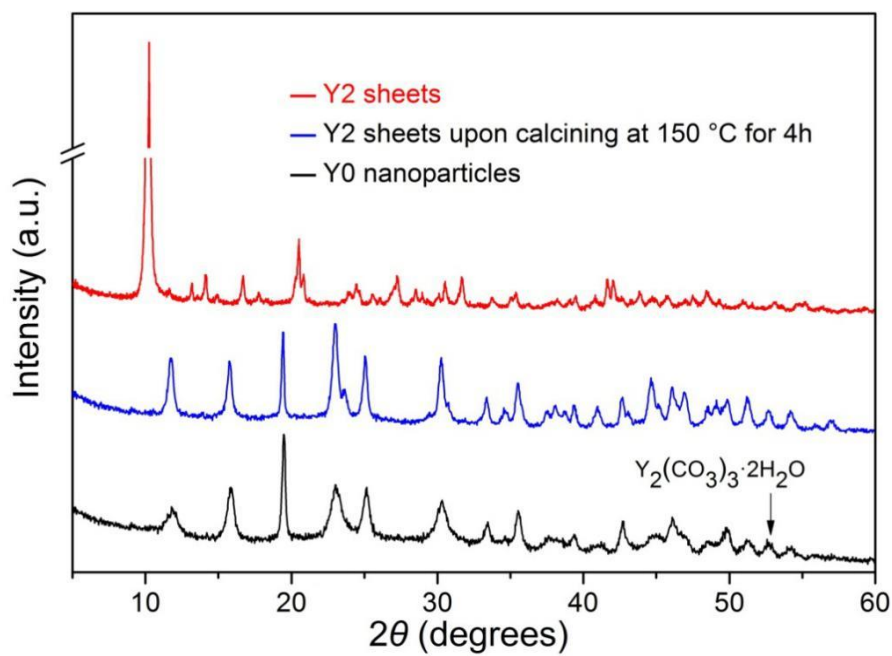

**Figure S19.** Powder XRD patterns of the samples. The patterns show the decomposition of $(NH_4) \cdot Y(CO_3)_2 \cdot H_2O$ to $Y_2(CO_3)_3 \cdot 2H_2O$ upon calcination at 150 ℃.

**SI Table S1.** Composition analysis of Y2 micron-sized sheets synthesized with the increasing addition of $NH_4HCO_3$.

| $NH_4HCO_3/Y_2(CO_3)_3 \cdot 2H_2O$ (molar ratio) | N (wt. %) | C (wt. %) | H (wt. %) | Y (wt. %) | Composition |
|---|---|---|---|---|---|
| 5/1 | 4.65 | 8.38 | 3.47 | 33.89 | $(NH_4)_{0.89} \cdot Y(CO_3)_{1.86} (OH)_{0.17} \cdot 2.81H_2O$ |
| 15/1 | 4.52 | 8.33 | 3.41 | 32.45 | $(NH_4)_{0.89} \cdot Y(CO_3)_{1.91} (OH)_{0.07} \cdot 2.93H_2O$ |
| 20/1 | 4.61 | 8.39 | 3.48 | 32.47 | $(NH_4)_{0.90} \cdot Y(CO_3)_{1.91} (OH)_{0.08} \cdot 2.91H_2O$ |

**SI Text**

# I.     Experimental Part

## A. FTIR analysis of Y2 micron-sized sheets

By comparing the FTIR spectra of $Y_2(CO_3)_3 \cdot 2H_2O$ nanoparticles and Y2 micron-sized sheets (**Figure S17**), one can see that besides the common absorption peaks Y2 possesses particular broad peaks at 2100-2150 cm$^{-1}$, clear peaks at 489 cm$^{-1}$ and 3608 cm$^{-1}$, and a few slight peaks as marked in **Figure S17**, indicating the presence of $NH_3$ and $H_2O$.[36-38] Especially, the peak at 1340 cm$^{-1}$ is characteristic of a hydrated yttrium ammine carbonate (Dolphin 1977, cited in ref.[39]). The above peaks disappear (**Figure S17**) after the calcination of Y2 at 150 °C for 4h, indicating the escape of ammonia coming from decomposition of Y2 and verifying the existence of $NH_4^+$ in Y2.

## B. Chemical composition analysis of Y2 micron-sized sheets

The element analyses show that the chemical composition of Y2 is $(NH_4)_x \cdot Y(CO_3)_y \cdot (OH)_z \cdot nH_2O$ ($0 < x < 1$, $1.5 < y < 2$, $0 < z < 1$, $x + 3 = 2y + z$, as listed in Table S1), which is hardly affected by the molar ratio of total $NH_4HCO_3$ to $Y(NO_3)_3$ when the ratio is greater than 5/1. Previous researches have indicated that the precise chemical composition of well crystallized Y2 is $(NH_4) \cdot Y(CO_3)_2 \cdot H_2O$, and the composition change of Y2 is due to the hydrolysis of $(NH_4) \cdot Y(CO_3)_2 \cdot H_2O$ in the wash cycles, as follows:[40, 41]

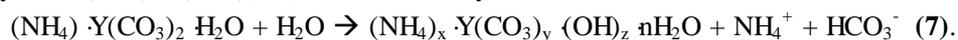

$(NH_4) \cdot Y(CO_3)_2 \cdot H_2O + H_2O \rightarrow (NH_4)_x \cdot Y(CO_3)_y \cdot (OH)_z \cdot nH_2O + NH_4^+ + HCO_3^-$  **(7)**.

In spite of the slight differences in chemical composition (*SI Appendix* **Table S1**) of Y2 arising from its hydrolysis, all of the XRD patterns of the products are identical within experimental errors (**Figure S18**), indicating that there is no apparent changes in the crystal structure, in line with the results reported in the hydrous yttrium carbonate mineral.[17]

## C. XRD analysis of Y2 micron-sized sheets

There are no crystallographic data of $(NH_4) \cdot Y(CO_3)_2 \cdot H_2O$ in the JCPDS database or reported in previous studies. On the other hand, due to the small size and the preferred orientation of Y2 sheets it is difficult to determine its crystal structure experimentally. Nevertheless, since all of the yttrium carbonate double salts including adamsite-(Y) (ideally $NaY(CO_3)_2 \cdot 6H_2O$), shomiokite-(Y) (ideally $Na_3Y(CO_3)_3 \cdot 3H_2O$), lecoqite-(Y) (ideally $Na_3Y(CO_3)_3 \cdot 6H_2O$), and a newfound rare earth double carbonates $(NH_4) \cdot Gd(CO_3)_2 \cdot H_2O$ are all layered structure consisting of both $YO_9/GdO_9$ polyhedra layers and $Na(CO_3) \cdot H_2O/NH_4 \cdot (CO_3) \cdot H_2O$ layers.[18, 42-44] Hence, the inference that Y2 (ideally $(NH_4) \cdot Y(CO_3)_2 \cdot H_2O$) possesses layered structure with the ammonium remaining in the interlamellar space is reasonable. In fact, Y2 can be considered to be $(NH_4) \cdot Gd(CO_3)_2 \cdot H_2O$ in which gadolinium is replaced by yttrium, and shows similar properties to $(NH_4) \cdot Gd(CO_3)_2 \cdot H_2O$. For instance, $(NH_4) \cdot Gd(CO_3)_2 \cdot H_2O$ is stable until 175 °C;[43] likewise, after calcination at 150 °C Y2 is decomposed to $Y_2(CO_3)_3 \cdot 2H_2O$ (**Figure S19**).

Hence, it can be concluded that Y2 is isostructural to $(NH_4) \cdot Gd(CO_3)_2 \cdot H_2O$, and possesses layered structure composed of $YO_9$ polyhedra layers and $NH_4 \cdot (CO_3) \cdot H_2O$ layers.

# II.     Theoretical Part

## A.  Calculation of the Equilibrium Concentration of $Y^{3+}$.

1. The Equilibrium Concentration of $CO_3^{2-}$

From the preparation procedure (Methods Summary part) the initial concentration of $NaHCO_3$ ($NH_4HCO_3$) is estimated to ~ 1/3 mol/L (i.e., 1/3 M), then the concentration of $HCO_3^-$ is ~ 1/3 M, due to the fact that $NaHCO_3$ ($NH_4HCO_3$) is strong electrolyte. The group $HCO_3^-$ is partly dissociated into $H^+$ and $CO_3^{2-}$ with an equilibrium constant of $K \sim 4.7 \times 10^{-11}$ at 25 ℃,[19] it follows that $K = [H^+] \times [CO_3^{2-}]/[HCO_3^-] = [CO_3^{2-}]^2/[HCO_3^-]$. Then, it can be deduced that the equilibrium concentration of $CO_3^{2-}$, $[CO_3^{2-}] = 3.9 \times 10^{-6}$ M.

2. The Equilibrium Concentration of $Y^{3+}$

With the equilibrium concentration of $CO_3^{2-}$ and the solubility product of $Y_2(CO_3)_3$ ($K_{sp} = 1.03 \times 10^{-31}$),[19] one has $K_{sp} = [Y^{3+}]^2 \times [CO_3^{2-}]^3$, and consequently $[Y^{3+}] = 4.1 \times 10^{-8}$ M. This is the equilibrium concentration of $Y^{3+}$ in the solution, before the onset of intercalation reaction.

**B. Calculation of Rate Constant for a Multistep Diffusion.** The diffusion of atoms from one atomic configuration to another is equivalent to a chemical reaction in the sense of the breaking of old bonds and the formation of new bonds. The speed of diffusion can be similarly describe by the rate constant: $K = Ae^{-E_a/(k_BT)}$. A multistep diffusion is just similar to a multistep chemical reaction. Consider the multistep diffusion of particle A from the original configuration (labeled by "0") via a number of intermediate configurations (labeled by "1, 2, 3,…, n-1") to the final configuration (labeled by "n") as shown in **Figure S13**.

The rate constant is $k_1$ for the diffusion from configuration 0 to 1, and is $k_2$ for the diffusion from configuration 1 to 2, and so forth (**Figure S13**). Given that the amount of substance A is $n_A$, then the time needed for the diffusion from atomic configuration 0 to atomic configuration $n$ is:

$t = \dfrac{n_A}{k_1} + \dfrac{n_A}{k_2} + \dfrac{n_A}{k_3} + ... + \dfrac{n_A}{k_n} = \sum_{i=1}^{n} \dfrac{n_A}{k_i}$. The average rate constant is $k = \dfrac{n_A}{t}$, which satisfies the relation: $\dfrac{1}{k} = \sum_{i=1}^{n} \dfrac{1}{k_i}$. It follows that the step with the smallest $k_i$ (equivalently, the highest activation energy $E_a$) plays a major role in determining the value of $k$, which is the so-called rate-determining step.[45]

**C. Estimation of the Debye length in the aqueous solution**

1. Concentration of the ions after the synthesis of $Y_2(CO_3)_3 \cdot 2H_2O$ nanoparticles

Upon the mixing of reactants ($1M \times 10ml$ $NH_4HCO_3$ drips into $0.2M \times 20ml$ $Y(NO_3)_3$, see Methods section), the synthesis reaction proceeds as follows:

$3NH_4HCO_3 + 2Y(NO_3)_3 + 2H_2O \rightarrow Y_2(CO_3)_3 \cdot 2H_2O + 3NH_4^+ + 3H^+ + 6NO_3^-$ **(C1)**

After the completion of reaction (C1), the amount of $H^+$ becomes $0.2\ M \times 20ml \times 3/2 = 0.006$ mol.

The amount of $HCO_3^-$ residue is $1M \times 10ml - 0.2\ M \times 20ml \times 3/2 = 0.004$ mol.

The $H^+$ and $HCO_3^-$ residues will react as follows:

$H^+ + HCO_3^- \rightarrow H_2CO_3 \leftrightarrow CO_2 + H_2O$ **(C2)**

After the completion of reaction (C2), the $HCO_3^-$ groups are depleted; the amount of residual $H^+$, $NH_4^+$ and $NO_3^-$ is $2M \times 1ml$, $1M \times 10ml$, $0.6M \times 20ml$, respectively.

To synthesize $(NH_4)Y(CO_3)_2 \cdot H_2O$, $1M \times 15ml$ $NH_4HCO_3$ is added to the solution. Considering the fast motions of $H^+$ in solution (due to its small mass), the newly added $HCO_3^-$ will be partly consumed by reaction (C2), while the $H^+$ will be nearly depleted.

2.  Calculation of the Debye length

At the initial stage of synthesizing the double-carbonates, the amount of the major residual solutes is as follows:

$n\_HCO_3^- = 1M \times 13ml$
$n\_NH_4^+ = 1M \times 10ml + 1M \times 15ml = 1M \times 25ml$
$n\_NO_3^- = 0.6M \times 20ml$

The total volume of the solution V = (20 + 10 + 15) ml = 45 ml.

Consequently, the concentration of the major residual solutes is as follows:

$c_1 = [HCO_3^-] = 1M \times 13ml/45$ ml = 13/45 M;
$c_2 = [NH_4^+] = 1M \times 25ml/45$ ml = 25/45 = 5/9 M;
$c_3 = [NO_3^-] = 0.6M \times 20ml/45$ ml = 12/45 M = 4/15 M.

Then the Debye length which describes the effective electrostatic interaction between the solutes can be calculated as follows:[46]

$\lambda_D = \sqrt{\frac{\varepsilon k_B T}{\sum_j n_j q_j^2}}$, where $\varepsilon$ is the dielectric constant, $k_B$ is the Boltzmann constant, $T$ is the temperature, $n_j$ and $q_j$ are respectively, the number of the $j$th solutes species and their charges. The quantity $n_j$ can be replaced with the molar concentration $c_j$ by multiplying the Avogadro number $N_A$.

Generally, the true concentrations of the freely moving ions are smaller than the values expected from the ideal solution, due the ion-ion interactions. As an approximation, the *activity coefficient* $\gamma$ is a good quantity to account for such a difference. The Debye length turns out to be: $\lambda_D = \sqrt{\frac{\varepsilon k_B T}{N_A \sum_j \gamma_j c_j q_j^2}}$.

In addition to the concentrations $c_1$, $c_2$, $c_3$ obtained above, we take $\gamma_1 = \gamma_3 = 0.7$ for $HCO_3^-$ and $NO_3^-$, and $\gamma_2 = 0.6$ for $NH_4^+$.[47,48] The dielectric constant $\varepsilon = 70\varepsilon_0$ with $\varepsilon_0$ being the permittivity of free space. The temperature $T = 333$ K. These parameters give that $\lambda_D = 5.09$ Å.

# SI References